# Structure and photoinduced effects in elemental chalcogens:
# A review on Raman scattering

Spyros N. Yannopoulos

*Foundation for Research and Technology Hellas – Institute of Chemical Engineering Sciences*

*(FORTH/ICE-HT), P.O. Box 1414, GR-26504, Rio-Patras, Greece*

## Abstract

Much progress has been made over a long period, spanning more than a century, in understanding the atomic arrangement on various length scales of non-crystalline chalcogens and their transitions upon certain external stimuli. However, it is broadly admitted that there are still several unsettled issues that call for proper rationalization. The current review presents an assessment of Raman scattering studies of non-crystalline phases of elemental chalcogens and their mixtures. First, a few remarks on the analysis of Raman data, related to polarization details and spectra reduction are presented. The effect of temperature, pressure and irradiation on the structure of chalcogens is reviewed in detail. As only selenium can form a stable glass at ambient conditions, the interest on sulfur and tellurium has been placed in the melt and the amorphous phase, respectively, whereas reference is also made to the sporadic structural studies of glassy sulfur at low temperatures. It is shown how Raman scattering can be exploited to explore unique phenomena emerging in the liquid state of sulfur, offering valuable information on the details of λ−transition including various thermodynamic-related properties. The subtle nature of this transition in selenium is also discussed. Tellurium is not only impossible to be prepared in the bulk glassy state, but also forms a very liable to crystallization amorphous film. Therefore, the emphasis is placed on light-induced nanostructuring and effects related to photo-amorphization and photo-oxidation.

## 1. Introduction

Non-crystalline chalcogenides, either as bulk glasses or amorphous films, have attracted extensive interest since their discovery, not only from the viewpoint of fundamental science but in addition owing to the technological potential that these materials exhibit over diverse sectors of active and passive applications [1,2]. Structure on the atomic scale is the decisive factor that determines macroscopic properties and ultimately function of the material. In addition, photoinduced effects, a hallmark of non-crystalline chalcogenides, depend on composition since structural arrangement must be suitable to allow photoinduced transformations on various length scales [3,4]. Understanding structure will not only allow identifying why a material behaves macroscopically in a certain manner, but will make feasible to establish structure-property relations. The latter are central to progress from serendipitous discoveries of functional materials to a rational design of materials with tailored properties. Disorder abolishes all the amenities accompanying the crystal periodicity, which render the role of structure-property relationships indispensable to disordered media.

Resolving structure in the amorphous state is *per se* a formidable task. This is true even for simple systems, e.g. monoatomic chalcogens, let alone for multicomponent materials. Research for applications in non-crystalline chalcogenides has since the beginning of their discovery moved to studies of multicomponent systems, i.e. ternary and quaternary glasses and amorphous films [5,6]. Looking over the extended archival literature of non-crystalline chalcogenides, it can be realized in several cases a deficit in our current understanding of structure on various length scales, despite the enormous volume of information accumulated. In certain cases, our weakness to understand structure in multicomponent chalcogenides derives from deficient knowledge of structure and stimuli-induced phenomena even for simple elemental chalcogens. This is true despite that elemental chalcogens have been in the focus much earlier than the discovery of amorphous chalcogenides. An early review on the utilization of Raman scattering in chalcogenides has been published by Ward [7]. Despite being monoatomic substances, elemental chalcogens exhibit a rich variety of structures, allotropes, and temperature-induced transitions which are reflected into the Raman spectra. Accomplishing a consistent and precise interpretation of the Raman spectra of elemental chalcogens is important in decoding the spectra of chalcogen-rich multicomponent glasses and amorphous films, where homonuclear chalcogen-chalcogen bonds abound and the formation of nanoscale chalcogen-based arrangements dominate the structure.

The scope of the current review is to present a comprehensive appraisal of the structure and photoinduced effects in elemental chalcogens. The content will primarily be focused on the use of Raman scattering to monitor and understand structure and effects in elemental chalcogens and mixtures of them under the application of external stimuli, such as temperature, pressure, illumination. Fallacies in the interpretation of Raman data where necessary will be reported. In selected cases reference to structure details handed by other techniques will be presented briefly when this is judged essential to supplement structural information obtained by Raman spectroscopy.

## 2. A glimpse at non-crystalline elemental chalcogens

Elemental sulfur is mostly known for the variety of its crystal phase and the spectacular phenomena observed in its molten state [8,9]. Indeed, much importance has been attached to crystalline sulfur since it exhibits the wider known variety of allotropic forms, at least 30, most of them being unbranched cyclic molecules. However, overwhelming attention has been drawn to the molten state of sulfur, in view of the extraordinary phenomena that the melt exhibits. The most celebrated among them is the thermoreversible polymerization transition where $S_8$ rings open to diradicals and concatenate to form long polymeric chains. The transition is known also as the (lambda) $\lambda-$transition owing to cusp-shaped singularity exhibited by the heat capacity resembling the capital Greek letter $\Lambda$. The density of the liquid also exhibits an inverse-shaped $\lambda$-transition at the same temperature. This transition is classified as a liquid-liquid phase transition (LLPT) where particular physical properties change abruptly or behave anomalously (contrary to common wisdom) around the transition temperature $T_\lambda \approx 159$ °C [10]. Several aspects of the liquid phase and the $\lambda-$transition have been compiled in previous reviews [8,9].

Experiments preparing glassy sulfur from the melt have already been conducted more than 150 years ago, as early reviews on the subject manifests [11]. Despite this, elemental glassy sulfur has received very limited attention, presumably because its technological potential is low compared to other chalcogenides. This essentially arises from the fact that its glass transition temperature $T_g$ – dependent upon the quenching temperature – is located well-below room temperature and the glass spontaneously crystallizes at ambient conditions. The low thermal stability of the glass has deprived scientists of exploring exciting aspects of this system, such as sub-$T_g$ structural rearrangements as well as other unique properties, since as this monoatomic glass can be prepared at a manifold of different structures based on the preparation conditions. It is edifying to mention at this point that regardless of the metastability of glassy sulfur and its high propensity to crystallization at ambient conditions, studying the glassy state is not just an experimental curiosity because the glass is a common state of sulfur in several planets and their satellites [12].

Glassy selenium exhibits a glass transition temperature which is located slightly above the ambient temperature, i.e. $T_g \approx$ 35-40 $^o$C, and the glass remains stable up to ~100 $^o$C. The thermal stability of the glass and its photoconductivity as a semiconductor, ~2.0 eV, justify the strong emphasis placed on this chalcogen [13,14]. It was therefore early recognized the vital role of non-crystalline selenium in a number of applications, including xerography, photovoltaics and X-ray detectors [1,15]. Undoubtedly, selenium has been the most thoroughly studied chalcogen in its non-crystalline state since it stands as an example of a monoatomic polymeric material. Despite exhaustive investigations, its structure is still a matter of research efforts to settle issues related e.g. to the relative concentration of rings-to-chains in the amorphous phase and the effect of ageing on the short- and medium-range structural order. A continuing source of controversy in interpreting Raman spectra of non-crystalline Se under the influence of various external stimuli has been the early-years misinterpretation of the main Raman bands assigned to $Se_8$ rings or chains, as will be discussed in Section 6.1.

In the way of sulfur, selenium also undergoes a temperature-induced polymerization transition where ring molecules open-up and join to form long chains [16]. However, the transition on these two elements presents an important difference. As discussed above, the $\lambda$–transition in sulfur takes place well above the melting temperature where a purely molecular liquid (composed mainly of $S_8$ rings) exists over a wide temperature regime. This makes easily accessible the polymerization transition by various experiments. An important structural difference between S and Se concerns the conformations of four-atom correlations, see Fig. 1. The *cis* conformation is more stable for Sulfur. Selenium, on the contrary, prefers the *trans* configuration which leads to chains as the energetically more stable structures. The consequence of this difference is that in all phases, the glass, the supercooled, and the melt, chains are the dominant species, which makes the polymerization equilibrium in Se subtle. A rough estimate has located the polymerization transition temperature of Se at ~356 K, which is much lower that the melting point of Se [16]. A recent survey on several properties on various phases of non-crystalline selenium can be found in a recent review [17].

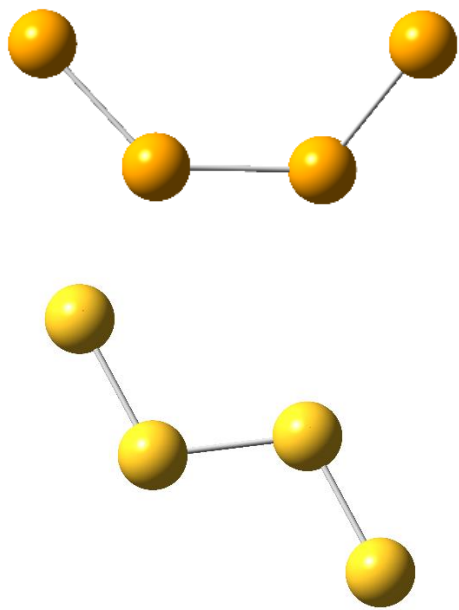

**Fig. 1:** Four-atoms conformations, (top) *cis*, leading to rings, and (bottom) *trans*, leading to chains.

Amorphous Te is the least explored chalcogen compared to S and Se, as it can be deposited on low-temperature substrates by vapor deposition or sputtering, whereas it speedily crystallizes if raised to room temperature. Bulk glassy Te cannot be produced, even if very high cooling rates are applied. Contrary to the other two chalcogens, no stable cyclic Te species are known today; tellurium structure is composed only of chains. Although structural studies of elemental amorphous Te appear scantily in the literature, the great preponderance of structural evidence is inferred from investigations of the liquid state by experiments and simulations. Notably, Te forms the weakest homonuclear bond in relation to S and Se. The bond energies for S-S, Se-Se and Te-Te bonds decreases in the order, 266, 192, and 149 kJ mol[-1] [18]. This is manifested in the higher sensitivity of Te to light, as it can photo-oxidize readily. While the element has received meager attention, its amorphous compounds, which exhibit important phase-change memory effects have almost monopolized research efforts over the last two decades [19].

### 3. A brief notice on structure of disordered media

A number of experimental techniques and various simulation tools have been employed to provide valuable information regarding structure, which, in selected paradigms, has been successful. However, often, structural information obtained by a particular experimental technique may contradict knowledge obtained by another one. This may arise from several causes. For instance, (a) different techniques probe diverse length scales, (b) data interpretation may neglect particular attributes of the experimental method used, (c) phase separation at the micro- or nano-scale has been neglected, and so on.

Unambiguously diffraction and inelastic scattering of X-rays and neutrons have handed us with crucial, albeit not complete, information about radial distribution functions and vibrational density of states (VDoS) in amorphous materials. Each technique is associated with certain advantages and weaknesses [3]. Other popular techniques that have been employed to probe structure are EXAFS, magnetic resonance, and photoelectron spectroscopy, as well as several others, which are out of the scope of the current review and will not be considered.

A useful attribute of disordered media is that clusters of atoms may be partially independent from their surroundings, which endows some degree of "molecularity" to these clusters. In other words, when the interactions among the atoms within a cluster are adequately stronger than the interactions with the neighboring atoms, this cluster is considered to assume a molecular character. This is encountered in elemental chalcogens and in most cases non-crystalline chalcogenides. Under such conditions, vibrational spectroscopy (Raman and IR) becomes a valuable tool as the details of the molecular clusters can easily and firmly be identified by exploiting *group theory* and *polarization analysis*. The decoupling of a vibrational unit from its surroundings can also happen in cases where their bridging angles to neighbors are close to the right angle. Elemental chalcogens convey a high degree of molecularity, which allows Raman scattering to remedy a gap left by diffraction techniques.

Raman spectroscopy is an operationally much easier experimental method compared to several other techniques. Excitation wavelengths can be used to straddle a very wide energy range moving from resonance to far-from-resonance conditions. Application of external stimuli, such as temperature and pressure, is often used as a means of retrieving additional information from disordered media. Raman spectroscopy is compatible with both high-temperature furnaces and high-pressure cells, especially because small scattering volumes of the material are required for the latter. Raman scattering has proven to be a very important tool in deciphering the atomic arrangement of disordered phases as it can provide information for both the short- and medium-range order. Besides, intramolecular mode frequencies are very sensitive to inter-molecular interactions. At this juncture, it is important to emphasize that the first-order Raman scattering of an amorphous medium conveys information of the entire phonon or VDoS. Care must be exercised, however, as the density of states is modulated by the light-to-phonon coupling coefficient or polarizability tensor elements.

Despite the above-mentioned advantages of Raman scattering can lead to ambiguous results or fallacies if not considered properly. On one side, the technique can be invasive to the material under study when the fluence is above the threshold initiating structural modifications, phase changes (e.g. crystallization), composition change (e.g. oxidation) or even decomposition of the material. On another front, it is not uncommon to see in several reports a risky approach attempting to identify Raman bands based on previous assignments of systems with the same atoms but different compositions, especially neglecting energetic factors of bond formation. Finally, it is important to emphasize that the structure of elemental chalcogens is typically low, i.e. 0-D for molecular systems such as molten sulfur below the $T_\lambda$, and 1-D for chains found in all three chalcogens under certain conditions. Therefore, interpretation of Raman data can be simplified by using this information, hence, extracting valuable information about the intra- and inter-molecular interactions.

## 4. Analysis of Raman spectra

Raman band intensities are frequently compared in order to draw conclusions about the relative species populations related to each band. The vibrational lines must be corrected to remove distortions engendered by the thermal population factor, eradicating also the effect of Rayleigh law scattering. The reduced representation, $I^{red}$, is therefore the spectrum that should be analyzed to achieve quantitative results. The reduced form is obtained by the experimental (raw) Raman spectra using the following relation [20]:

$$I^{red} = \frac{\omega}{(\omega_0 - \omega)^4} \left[ n(\omega, T) + 1 \right]^{-1} I_{exp} \propto C^{\alpha\beta}(\omega) \cdot g(\omega) \tag{1}$$

the first term stands for the correction for the wavelength dependence of the scattered intensity. It should be borne in mind that the reduced spectrum represent the VDoS modulated by the coupling coefficient (matrix elements). The term $[n(\omega, T) + 1]$ is the Bose-Einstein distribution function, at temperature $T$, for the Stokes-side Raman scattering, given by the equation:

$$n(\omega, T) = \left[\exp\left(\hbar\omega / kT\right) - 1\right]^{-1} \tag{2}$$

It is common in Raman studies of amorphous media (liquids or disordered solids) to employ polarization-dependent analysis of the scattered light, in order to retrieve crucial information on the symmetry of the vibrational modes and thus characterize and assign more reliably various vibrational modes. Typical polarization geometries applied experimentally are the polarized (VV or HH) and the depolarized (HV or VH) ones. VV denotes that the polarization direction of the incident and scattered radiation are both vertical, perpendicular to the scattering plane, while HV denotes cross polarization of the incident and scattered beams. Figure 2(a) displays typical polarized and depolarized Raman spectra of liquid sulfur at various temperatures. Therefore, the polarizability tensor is composed of isotropic and anisotropic components, which arise from the diagonal and off-diagonal elements of the polarizability tensor, respectively. These two components are extracted from the polarized ($I_{VV}$) and depolarized ($I_{HV}$) scattered intensities via the relations [21]:

$$I_{iso}(\omega) = I_{VV}(\omega) - \frac{4}{3}I_{HV}(\omega) \quad , \qquad I_{aniso}(\omega) = I_{HV}(\omega) \tag{3}$$

$I_{iso}$ carries information on the purely vibrational spectrum (totally symmetric vibrations). A characteristic isotropic spectrum of liquid sulfur at 160 ºC [22] is shown in the inset of Fig. 2(b), where the depolarized band at 150 cm$^{-1}$ has vanished. The anisotropic part contains only depolarized Raman intensity, which is strongly influenced by orientational motion and deformations of "molecular" units. The reduced representation (Eq. 1) is frequently applied to the $I_{iso}$ and $I_{aniso}$ components of the scattered intensity. Figure 2(b) contains the reduced $I_{iso}$ and $I_{aniso}$ representations of the spectra shown in the left panel. One should pay attention to the immense change in the relative intensity ratio among low and high frequency Raman bands; the high frequency bands intensify upon removing thermal population effects from the Raman spectra.

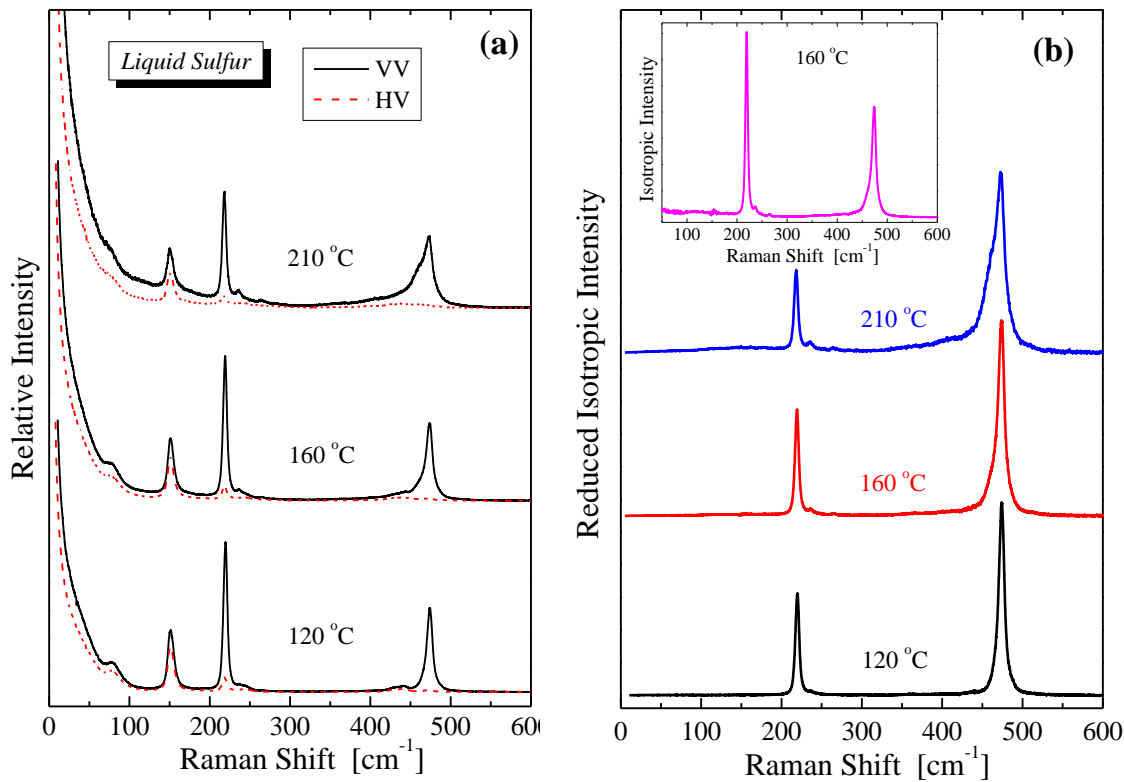

**Fig. 2:** **(a)** Representative polarized (VV) and depolarized (HV) Stokes-side Raman spectra of liquid sulfur at 120 ºC, 160 ºC, and 210 ºC. **(b)** Reduced isotropic representation of the same spectra. The inset shows the isotropic spectrum at 160 ºC before reduction with the thermal population factor, in order to reveal the change in the peak intensity ratio before and after the reduction. "Reprinted from [A. G. Kalampounias, K. S. Andrikopoulos, S. N. Yannopoulos, J. Chem. Phys. 118, 8460–8467 (2003)], with the permission of AIP Publishing."

## 5. Sulfur

### 5.1 Liquid Sulfur

Elemental sulfur is one of the most characteristic examples of a monoatomic solid that exhibits a dazzling variety of allotropic forms as it has been suggested to exist in more than 30 solid modifications [23]. Perhaps, the most fascinating feature of sulfur is the thermoreversible polymerization transition, at $T_\lambda \approx 159$ ºC [8,9], which belongs to the class of LLPTs. The structure of the low-temperature liquid, in the range from melting point to $T_\lambda$, is dominated by $S_8$ rings; other species may coexist at low concentrations. At $T_\lambda$, the living polymerization transition manifests itself via the occurrence of observed anomalies in the temperature dependence of various physical parameters. The transition involves two steps. The $S_8$ rings first open to form diradical $S_8^*$ chains (initiation step) and concatenate to form long chains $S_n$ (propagation step). This transition becomes visually evident by color change of the melt and is accompanied by a spectacular increase of viscosity by almost four orders of magnitude, from 0.007 Pa s at 157 °C to 92 Pa s at 187 °C [24]. Photon correlation spectroscopy was employed to provide evidence for the existence of a relaxation process in liquid sulfur at the kHz frequencies [25]. This finding provided the rationale underlying the $\lambda$−transition, ruling out a simple coupling between viscosity and structural relaxation in this system.

The extent of polymerization $\phi(T)$ expresses the relative weight fraction of polymer to monomer content as a function of temperature. Research commenced more than 170 years ago, recognized $\phi(T)$ as the most valuable parameter of sulfur's LLPT (living polymerization). Based on theoretical approaches [26,27,28,29] and computer simulations [30], $\phi(T)$ has been used to explore several thermodynamic quantities.

Raman spectroscopy has played an essential role in providing a means for a reliable, *in situ* estimation of the polymer content up to very high temperatures [22,31]. In this context, a long standing issue was tackled paving the way for investigations of the λ–transition of sulfur under the influence of various other conditions, as is discussed below. Indeed, on the experimental side $\phi(T)$ is determined by *ex situ* laborious experiments employing a procedure, quench-and-dissolution method, that often requires inextricable procedures and the adoption of severe approximations [32]. As Fig. 3 shows, $\phi(T)$ can be estimated by *in situ* temperature dependent Raman spectra, with high accuracy, by exploiting the symmetric S-S bond-stretching vibrational mode changes from ~472 cm$^{-1}$ for $S_8$ rings to ~461 cm$^{-1}$ for $S_n$ chains. Figure 3(a) illustrates the systematic change of the relative intensity ratio of the bands of the monomer and polymer. Figure 3(b) shows the temperature dependence of the intensity ratio $I^{S_n}/(I^{S_n} + I^{S_8})$ that corresponds to the polymeric content, as derived from fitting of the Raman spectra with Lorentzian lines. In this analysis it is tacitly assumed that both bands have comparable Raman matrix elements. This is not an unreasonable hypothesis since both bands are related to S−S bond stretching modes. Given that the matrix elements depend on the change of the polarizability during the vibrational motion, it suffices to compare the polarizability of the S−S bonds in *cis* and *trans* conformations. The polarizabilities of sulfur atoms participating in rings and chains have been estimated by dielectric constant measurements and were found to differ by less that ~7%) [33]. Besides, the modulation of the polarizability during a vibration depends on the geometrical details of the relevant bond, which is very similar for rings and chains. Therefore, what emerges from these arguments is that comparison of the Raman intensities of the above two bands can indeed be considered as a reliable indicator of the polymer content in liquid sulfur.

Data extracted from Raman scattering are compared with those obtained by the quench-and-dissolution method [32]. The two data sets qualitative agree up to some temperature above which they deviate. The difference may be due to various reasons. In particular, the quench-and-dissolution methods has the tendency to underestimate the polymer content. This arises from the finite quenching rate which cannot assure instant freezing of the liquid and preservation of its polymer content. Another reason is the metastability that polymer species exhibit when exposed to ambient conditions. The solid and dashed lines in Fig. 3(b) represent the predictions of theoretical models (see figure legend for description). It should be stressed here that Raman scattering can provide a wealth of thermodynamic-related information based on the description of the $\phi(T)$ data in terms of theoretical predictions. More details can be found elsewhere [34].

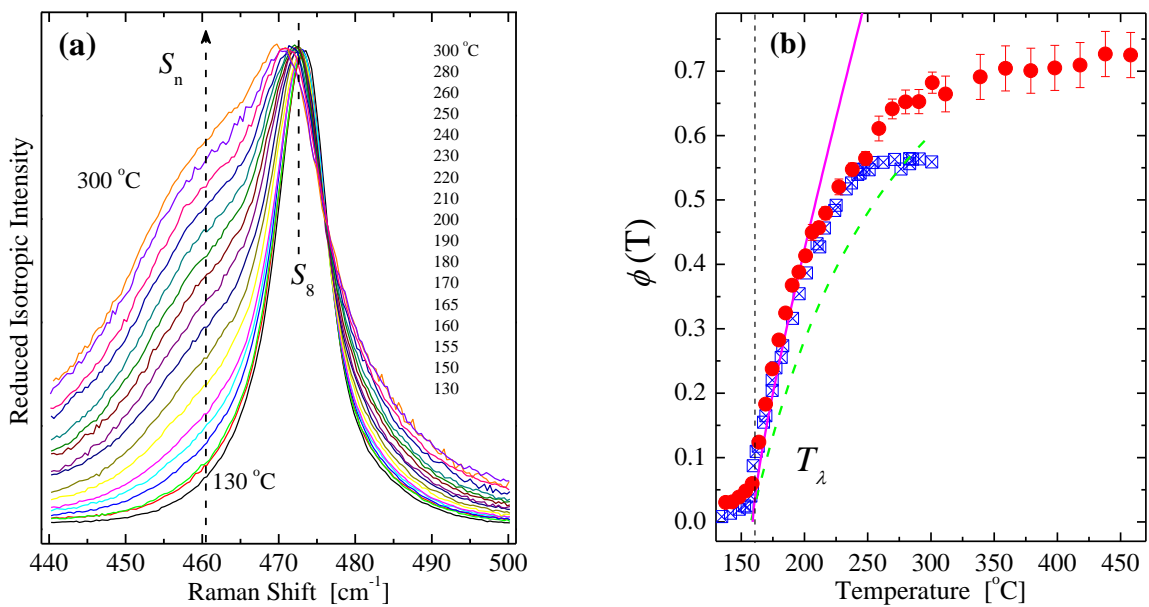

**Fig. 3: (a)** Reduced isotropic Raman spectra of liquid sulfur at various temperatures [22]. The dashed lines denote the positions of the bands assigned to the monomer ($S_8$) and polymer ($S_n$) species. **(b)** Temperature dependence of the extent of polymerization $\phi(T)$ represented by the intensity ratio $I^{S_n}/(I^{S_n} + I^{S_8})$ where $I^{S_n}$ and $I^{S_8}$ denote the integrated intensities of the symmetric bond-stretching vibrational mode (S-S) corresponding to polymeric and $S_8$ ring species, respectively [22,31]. Crossed squares: data from other methods [32]. The thick solid line represents the n-vector model prediction [28], while the dashed line the prediction of the Tobolsky-Eisenberg model [27]. The error bars are observable for those temperatures where they are larger than the size of the symbols used. "Reprinted from [A. G. Kalampounias, K. S. Andrikopoulos, S. N. Yannopoulos, J. Chem. Phys. 118, 8460–8467 (2003)], with the permission of AIP Publishing."

### 5.2 Glassy Sulfur

Apart for the unique LLPT, sulfur is notably one of the very few monoatomic substances that can be obtained in the bulk glassy state once quenched from any temperature $T_q > T_\lambda$ [8]. The presence of polymeric sulfur is the prerequisite for glass formation. The glass structure contains both types of molecules, i.e. $S_8$ and $S_n$. As discussed above, the $\lambda$–transition has been extensively studied and adequately understood. On the other hand, although being a long-standing issue, the attention on the elucidation of the structure of glassy sulfur has been sporadic and not fully conclusive [11,35,36,37,38,39,40].

Exploring the structure and properties of g-S is complicated *per se* as the glassy state exhibits also a kind of allotropy. Indeed, the glass structure and its properties, depend strongly upon preparation conditions and especially on the temperature where the melt is equilibrated before being quenched. This is a consequence of the temperature-dependent equilibrium between rings and chains in the melt. Quenching captures the equilibrated structure at a given temperature into a metastable solid configuration, which means that several glasses, differing in structure and properties, can be prepared by one substance. The metastability of the quenched product implies that when configurational degrees of freedom are unfrozen at $T > T_g$ kinetic effects concerning the rings$\leftrightarrow$chains equilibrium play a dominant role in structural correlations. The glass transition temperature increases with increasing polymer content. The $T_g$ of sulfur quenched from 200 ºC is estimated at $\approx$ –30 ºC, while purely polymeric Sulfur (*Crystex*) obtained after dissolving $S_8$ molecules, is a semi-crystalline product with $T_g \approx 70$ ºC [35].

It should be noted that stable bulk amorphous sulfur was obtained by using a recently developed rapid compression method from the melt using the following procedure [41]. First, sulfur crystal was pre-pressed to 0.17 GPa, and then melted at 423 K at this pressure. The melt was then solidified at 423 K through a rapid compression from 0.17 to 2 GPa in 20 ms, and finally cooled down naturally to room temperature and decompressed slowly to atmospheric pressure. More recently, deep supercooling of liquid sulfur by 150 ºC, i.e. up to –35 ºC, was achieved by direct electrochemical generation of cooled sulfur microdroplets from polysulphide electrochemical oxidation on various metal-containing electrodes [42].

Typical full range Raman spectra of glassy sulfur, quenched from 200 °C, at various polarization geometries and representative fitting results are shown in Fig. 4 [40]. The low frequency range of the Raman spectrum reveals the presence of the Boson peak at ~20 cm⁻¹, which is a genuine feature of non-crystalline media. A detailed temperature dependent study of the vibrational modes revealed that for torsional/librational modes thermal coefficients of the glass frequencies, $d\nu/dT$, are greater than those of the corresponding crystal by a factor of two. On the contrary, in the high wavenumber region, the thermal coefficients of the glass are lower than those of the crystal. Raman spectra revealed that quenching of molten sulfur (vacuum sealed in silica tubes) preserves reasonably well the polymer content into the glass. This is manifested in Fig. 5(a) by the comparison of the reduced isotropic Raman spectra of the melt (473 K, $\phi^{melt} \approx 0.45$) and the quenched product measured at 93 K ($\phi^{glass} \approx 0.42$). Raman spectra at such low temperatures, devoid of thermal-induced broadening effects, have made it possible to resolve two bands within the polymer species spectral range. *Ab initio* calculations were employed to simulate the vibrational frequencies, (see Fig. 5(a) for the theoretical spectrum) of polymeric chains $S_{8k}$ ($k = 1$, …, 7). These calculations provided a means of assigning two types of polymeric chains, i.e. short and long ones. Sulfur chains were found to vibrate at higher frequencies for chain length up to $k=2$ (or 16 S atoms); frequencies remain almost constant beyond that threshold. The combination of Raman data and *ab initio* calculations helped settling the question when a sulfur chain can be considered as a polymer or what is the critical value of $n_{cr}$ above which a chain behaves as a polymer from the vibrational point of view.

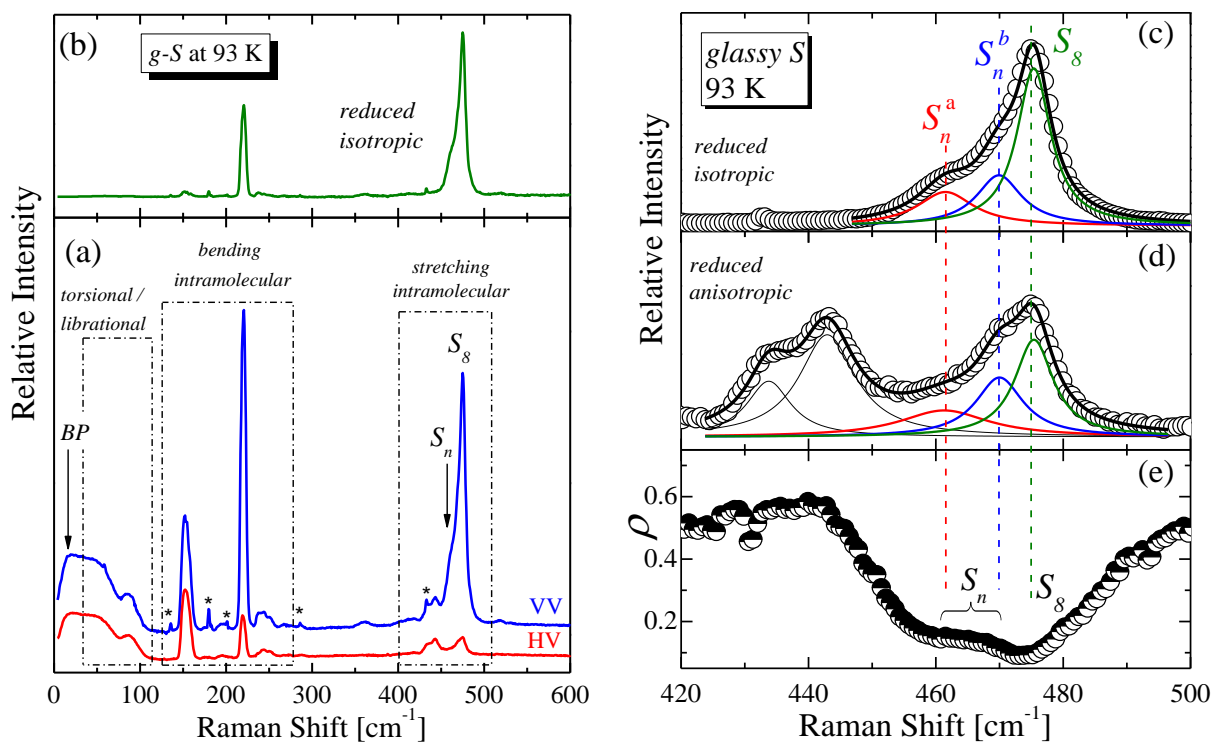

**Figure 4:** **(a)** Polarized, anisotropic and **(b)** isotropic Raman spectra of g-S at 93 K. Stars denote He-Ne laser plasma lines. Frequency ranges for the torsional/librational, bending/stretching intermolecular are also denoted. Representative fitting example of **(c)** isotropic and **(d)** anisotropic Raman spectra for the S-S stretching modes. **(e)** Frequency dependence of the depolarization ratio, $\rho$. Excitation wavelength: 632.8 nm; spectral resolution: 1.5 cm⁻¹. "Reprinted from [K. S. Andrikopoulos, A. Kalampounias, O. Falagara, S. N. Yannopoulos, J. Chem. Phys. 139, 124501 (2013)], with the permission of AIP Publishing."

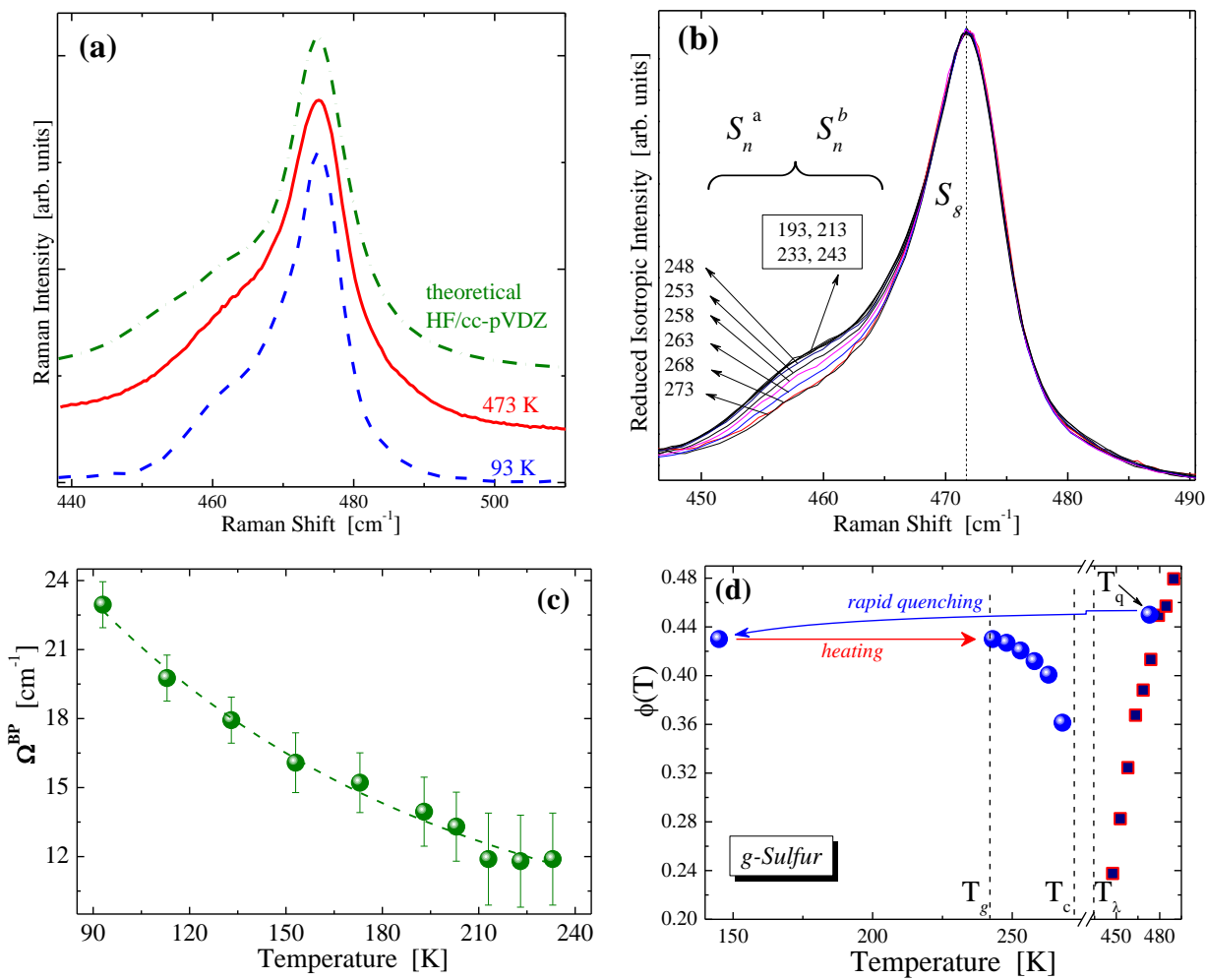

**Figure 5: (a)** Reduced isotropic intensity of the S-S bond stretching spectral region for the parent liquid (solid red curve) at 473 K and the quenched product (dashed blue curve) at 93 K. The upper spectrum (dashed-dotted green curve) is the theoretically calculated spectrum representing the vibrational modes of a mixture of short and long chains and $S_8$ units. **(b)** Intensity normalized Raman spectra of the S-S stretching modes at various temperatures. **(c)** "Phase diagram" of the polymer content of elemental Sulfur. Data at $T > T_\lambda$ are from Ref. [22]. The polymer content at $T_q$ is slightly higher than the polymer content of the quenched product. At temperatures below 243 K $\phi(T)$ maintains a constant value of ~0.43, while decrease of the polymer content commences above $T_g$. **(d)** Temperature dependence of the Boson peak maximum, $\Omega^{BP}$. "Reprinted from [K. S. Andrikopoulos, A. Kalampounias, O. Falagara, S. N. Yannopoulos, J. Chem. Phys. 139, 124501 (2013)], with the permission of AIP Publishing."

The interpretation of Raman spectra in g-S is even more perplexed because the glass structure is vividly changing upon heating at $T > T_g$, see Fig. 5(b). This arises from the metastability of the polymer chains which are thermodynamically forced to convert to the more stable $S_8$ rings. This causes a systematic reduction of the band intensity associated with $S_n$ chains. The changes take place at $T > 243$ K, which matches the $T_g$ of g-S. The sub-$T_g$ chain-to-ring transformation is reminiscent of a secondary relaxation process, which could also responsible for the surprisingly strong softening of the Boson peak frequency, Fig. 5(c). The latter typically exhibits red-shift as long as the glass is heated at $T > T_g$, while glassy sulfur appears as an exception to this rule exhibiting strong sub-$T_g$ softening. A secondary relaxation process was proposed long ago [43] in studies of the mechanical relaxation properties of g-S and S-rich binary As-S alloys. To account for the temperature dependence of viscosity, a process involving bond interchange, beyond normal diffusional motion, was suggested. Therefore, Raman data have provided evidence on the origin of this less understood process as the bond interchange can be related to the $S_8 \leftrightarrow S_n$ transformation. Figure 5(d) compiles data for the polymer content $\phi(T)$ over the widest accessible range where the material exist in non-crystalline form including the glass, the supercooled liquid and the melt.

### 5.3 $\lambda$−transition of liquid sulfur under nanoscale confinement

A number of experiments has shown that when matter is confined to pores with dimensions limited to a few molecular diameters modifications of characteristic features associated with phase transitions are expected. When it comes to polymerization transitions, geometrical confinement serves as a means that impedes the free growth of the polymeric species. The study of living polymerization processes into restricted geometries has many implications at various sectors such as the synthesis of polymers with unusual properties in sub-micrometer reactors. In addition, it offers a means to investigate changes in protein folding in confined spaces, which mimics the crowded environment of the living cell. Based on the methodology discussed in the previous sections, Raman scattering has proven a powerful tool for studying the behavior of sulfur's LLPT under confinement in nanoscale pores [44,45]. Liquid sulfur was confined by allowing the melt to be infiltrated into cylindrically–shaped (5 mm diameter, 3 mm height) porous silica matrices of three different pore diameters, i.e. 2.5, 7.5, and 20 nm, sealed under vacuum in silica tubes. The effect of confinement on the $\lambda$–transition of sulfur under confinement is illustrated in Fig. 6(a) for selected temperatures. The difference between the spectra in the range of the vibrational modes of the polymeric content becomes more prominent at higher temperatures. The effect is more severe for the liquid confined in the 2.5 nm diameter pores. A quantitative description of this effect is shown in Fig. 6(b), which illustrates the temperature dependence of the polymer fraction, $\phi(T)$. The corresponding curves of the bulk liquid and the liquid confined within the 20 nm pores are practically the same. Both curves exhibit steep rise at $T_\lambda$ and the polymer content is similar for all temperatures. Confinement of the liquid into the 7.5 and 2.5 nm pore-size matrices gradually moderates the $\phi(T)$ data ate elevated temperatures. In particular, $T_\lambda$ shifts to higher temperature, the rate of increase of $\phi(T)$ becomes smaller and the extent of polymerization decreases. The polymerization transition of sulfur is a two-step process [27] described by the reactions, the initiation step (scission of the $S_8$ rings to $S_8^*$ diradicals) and the propagation step (concatenation of the $S_8^*$ diradicals). Restriction to nanoscale pores affects the second step causing the changes in $\phi(T)$ curves described above.

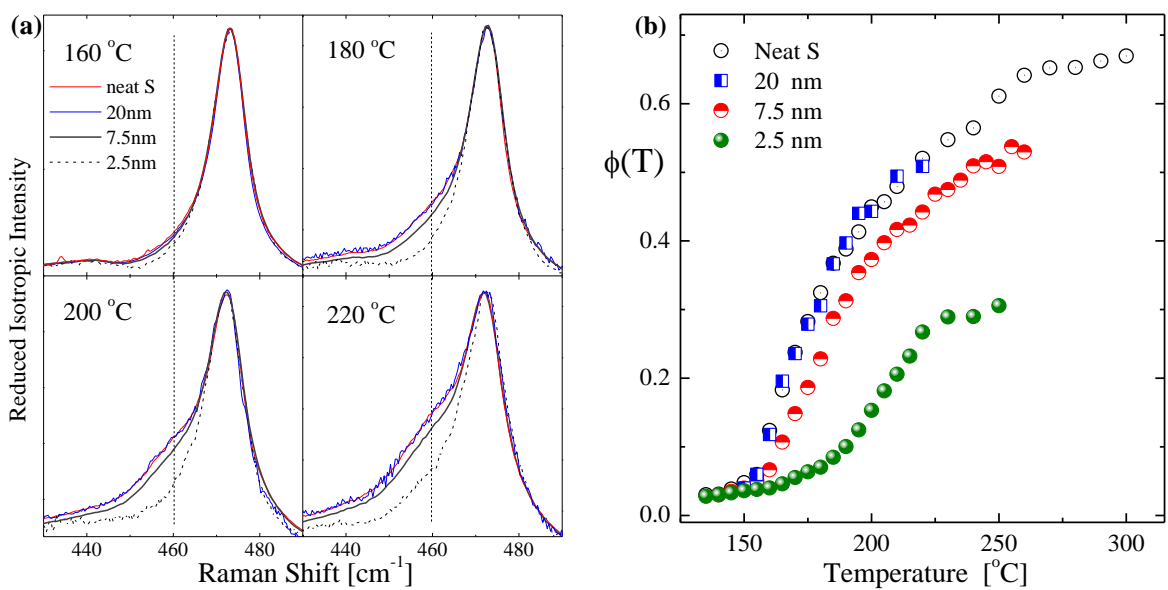

**Figure 6:** **(a)** Reduced isotropic Raman spectra of bulk and confined sulfur at various temperatures and pore sizes. The spectra are normalized with respect to the $S_8$ peak at ~470 cm$^{-1}$. **(b)** Temperature dependence of the polymer fraction, $\phi(T)$, for bulk and confined sulfur. [K. S. Andrikopoulos, A. G. Kalampounias, S. N. Yannopoulos, Soft Matter 7, 3404–3411 (2011)] - Reproduced by permission of The Royal Society of Chemistry.

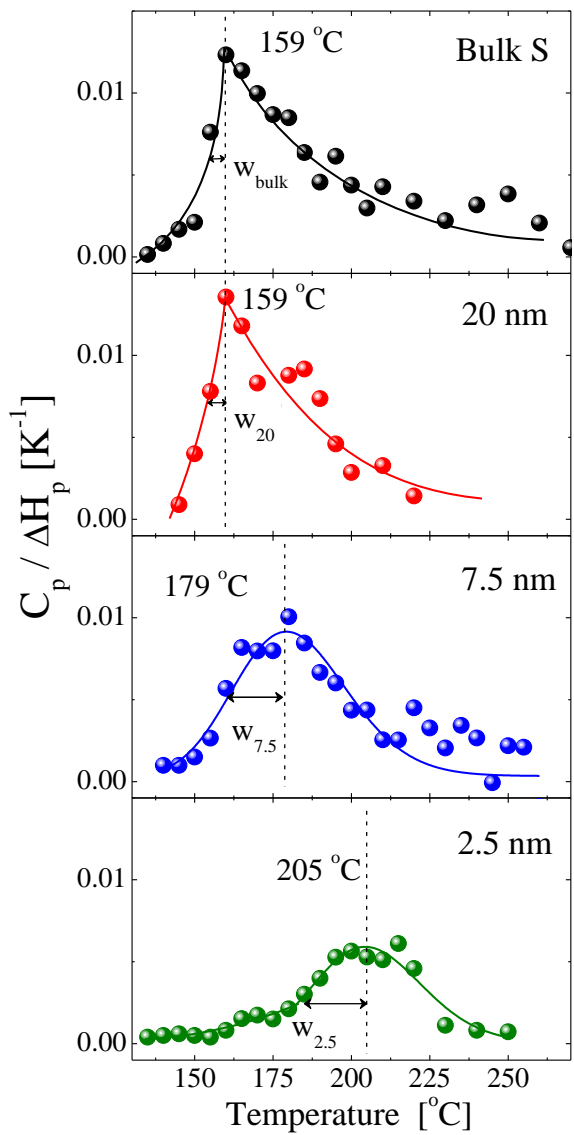

**Figure 7:** Heat capacities for bulk and confined sulfur. The vertical dashed lines and the labels denote the corresponding polymerization transition temperatures. [K. S. Andrikopoulos, A. G. Kalampounias, S. N. Yannopoulos, Soft Matter 7, 3404–3411 (2011)] - Reproduced by permission of The Royal Society of Chemistry.

The unprecedented behavior of liquid sulfur has intrigued theorists. A nonmean field model for polymerization transitions developed by Wheeler and co-workers [28,46] has provided useful relation to understand the changes in the sharpness of the living polymerization transition due to external parameters (i.e. confinement). Uniquely, the theory provides a means to calculate the constant-pressure heat capacity, $C_p$ through $\phi(T)$ values obtained by Raman scattering. This can be achieved using the relation:

$$C_p(T) \approx \Delta H_{\text{prop}} \frac{d\phi(T)}{dT}.$$ (4)

$\Delta H_{\text{prop}}$ stands for the enthalpy of the propagation step in the polymerization reaction. The heat capacities for the bulk and the confined sulfur have been calculated using the above relation, see Fig. 7.

While the transition for the liquid confined in 20 nm pores is as sharp as that of the bulk liquid, the loss of the features of a second-order transition emerges for more severe confinement (2.5 and 7.5 nm) accompanied by appreciable rounding effects. Changes in heat capacity pertain to: (a) the magnitude (height), (b) the temperature of the maximum, and (c) the asymmetry of the curves (rounding), in comparison to the bulk sulfur curve. For more details about the information that can be gained by exploring changes in heat capacities, see Ref. [45]. Extending the model of sulfur polymerization transition to account for the confinement entropy of the chains and rings, Simon and co-workers [47] have provided analytical and numerical solutions that fit the experimental data of Fig. 6(b) on a quantitative level.

### 5.4 The role of impurities on Sulfur's $\lambda-$transition

The polymerization transition of Sulfur in the case of intentional "impurities", that is, when alloyed with small (< 5 at. %) concentrations of other elements, such as Se and As has also been studied into detail [48]. Theoretical approaches have provided a solid basis for understanding changes in the kinetics and thermodynamics of such effects [29, 49] expressed as "rounding" effects which are substantiated as a smearing of the sharp changes accompanying the temperature dependence of various physical properties around a second-order-type or $\lambda-$transition point. Such theories have elaborated the case of initiators and the effect they engender in the polymerization reaction. Initiators are molecules usually incorporated into the monomeric phase to commence and facilitate polymerization. In the paradigm of Sulfur, heteroatoms such as Se and As can play such role. However, in that case, Se and As do not only play the role of the initiator of the polymerization reaction, but also they participate in the structure of the polymerized melt. Early studies combining Raman spectroscopy and differential scanning calorimetry by Ward [50, 51] showed that $T_\lambda$ exhibits systematic decrease upon the addition of Se and As in Sulfur. The drop of $T_\lambda$ caused by As is abrupt, i.e. ~40 °C for 2 at.%, which remains almost unchanged for higher concentration, while Se causes a rather linear dropt of $T_\lambda$ reaching ~110 °C for 15 at.% Se [51].

More detailed studies using Raman spectroscopy and analyzing the spectra in the context described in Section 3.1 were conducted by Andrikopoulos *et al.* [48]. Representative Raman spectra of dopes sulfur at various spectral ranges are shown in Fig. 8.

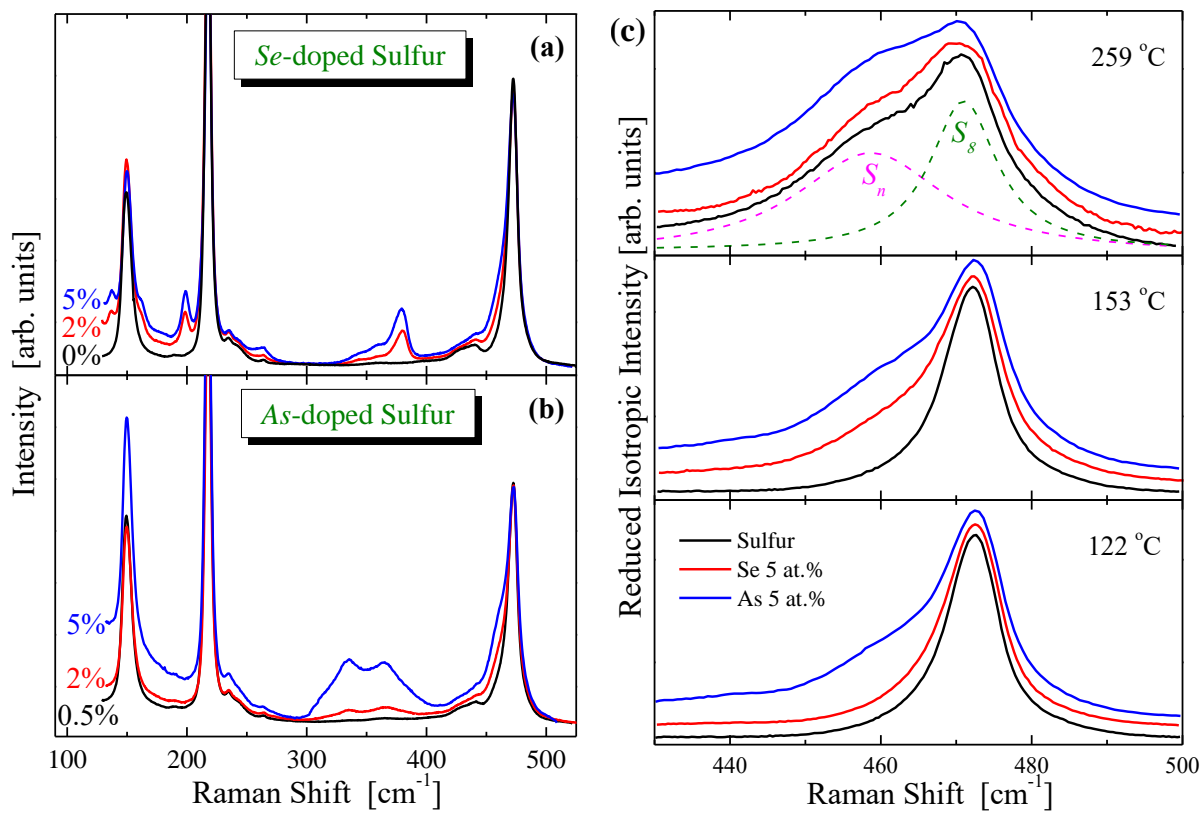

**Fig. 8:** Raman spectra recorded at 143 °C of **(a)** Se-doped sulfur (2% and 5%) and neat sulfur and **(b)** As-doped sulfur (0.5%, 2%, and 5%). **(c)** Reduced isotropic Raman spectra of the S···S stretching spectral region recorded at 122, 153 and 259 °C of neat sulfur and sulfur mixed with 5 at.% of Se and As. The dashed lines represent the components of monomers and polymeric content. "Reprinted (figures 1, 2) with permission from [K. S. Andrikopoulos, A. G. Kalampounias, S. N. Yannopoulos, Phys. Rev. B 72, 014203 (2005).] Copyright (2005) by the American Physical Society."

Extracting the fraction of S atoms residing in polymeric chains, $\phi(T)$, as discussed above, Fig. 9(a), has revealed qualitative and quantitative differences between Se and As-doped sulfur melts. The main features of the $\lambda$-transition of the doped sulfur in relation to the neat one are: (i) The $\phi(T)$ curves shift to lower temperature, (ii) the steepness of the rise of the polymer content curve vs. temperature is moderated proportionally to the dopant content, (iii) the saturation (plateau) value of the $\phi(T)$ curves are slightly higher compared to the corresponding ones of neat sulfur [48]. The role of As atoms is vital in weakening the S–S bond, lowering the energy required for the scission of $S_8$ rings, which takes place at temperatures lower than neat sulfur's $T_\lambda$.

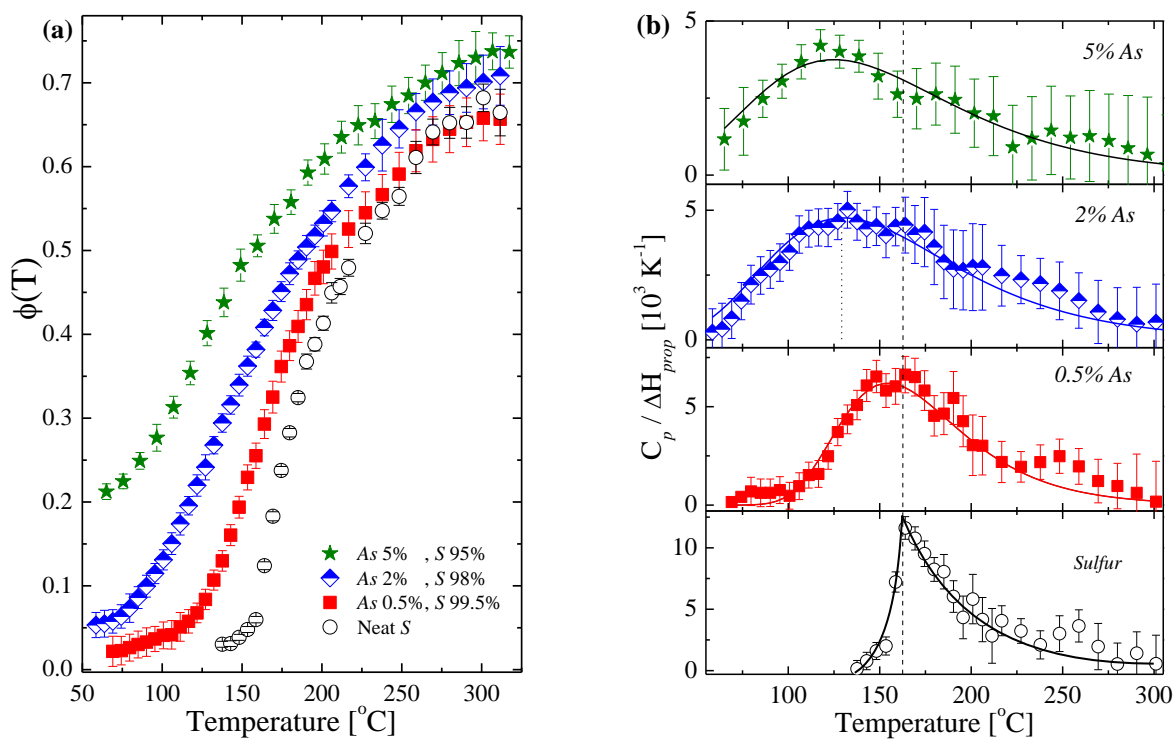

**Fig. 9:** **(a)** Temperature dependence of the polymer content for sulfur melts doped with various *As* concentrations. **(b)** Heat capacities for bulk and doped sulfur calculated form the data shown in (a). "Reprinted (figures 4, 5) with permission from [K. S. Andrikopoulos, A. G. Kalampounias, S. N. Yannopoulos, Phys. Rev. B 72, 014203 (2005).] Copyright (2005) by the American Physical Society."

The data of the polymer content $\phi(T)$ can be transformed to a useful thermodynamic variable using Eq. (4). A parameter proportional to the heat capacity $C_p(T)$ is shown in Fig. 9(b). $C_p(T)$ is sensitive to the changes that the sharpness of the polymerization transition experiences. The data show that even at very low concentrations of As (0.5 at. %) the $\lambda$-transition undergoes significant rounding or loss of sharpness and shifts to lower temperature. Rounding becomes progressively noticeable at elevated Arsenic doping concentrations and becomes evident as a decrease of the specific heat maximum and an increase of the width of the $C_p(T)$ peak. Similar analysis of the Se-doped sulfur showed that melts with 2 and 5 at. % Se exhibit less significant changes in comparison to analogous percentages of As. Even more interesting is the finding that although the presence of Se stimulates the polymerization transition at temperatures below that of neat sulfur, the sharpness of the transition is not affected as it is observed for doping with As.

### 6. Selenium

As has already been stated, selenium is the only chalcogen that can be prepared and can remain as a stable glass at ambient conditions. This attribute of Se and its numerous high-tech applications into various sectors, such as xerography, detectors for medical imaging, and so on, have been the incentive for a vast number of investigations [4,13,14]. Selenium, just like sulfur, also exhibits a wide variety of stable molecules. The crystalline allotrope of trigonal (grey or metallic) selenium, t-Se, is the thermodynamically most stable than other allotropes. t-Se is composed of helical chains with a repeat unit of three atoms, which are arranged in parallel, leading to trigonal symmetry. Covalent bonds determine intrachain interactions whereas the weaker van der Waals forces dictate the interchain interactions. As the intermolecular distances are shorter than the der Waals distances, it is deduced that some weak attraction of covalent character exists between neighboring chains. The partial overlap of the unoccupied lone-pair orbital in one Se atom with the unoccupied $p$-like antibonding orbital in a neighboring atom causes this weak interchain interaction. The monoclinic form of Se is found in three different allotropes, the $\alpha$-, $\beta$-, and $\gamma$-Se, each one composed of eight-membered $Se_8$ rings.

In analogy with sulfur, selenium also undergoes a thermoreversible temperature-induced polymerization transition [16], albeit the transition for the two chalcogens differs significantly. The $\lambda-$transition of sulfur occurs in the melt; hence, between the melting point of the crystal and the polymerization temperature the liquid is composed, almost exclusively, of $S_8$ rings. This renders the transition sharp and easily observable. In selenium, atoms have higher tendency, in comparison to sulfur, to be disposed in the *trans* configuration, which leads to the formation of chainlike structures. As a result, none of the non-crystalline phases of Se (glass, supercooled, melt) is composed entirely of $Se_8$ rings. Despite that several research efforts have been focused on the elucidation of the nature of the polymerization transition and the transition temperature $T_\lambda$ in Se, these issues are still vividly debated.

### 6.1 Glassy and amorphous Selenium

Non-crystalline selenium is undoubtedly one of the most explored substances among other chalcogens and chalcogenides. Studies include the bulk glass and thin films prepared by evaporation, sputtering and pulsed laser deposition). Its structure has been studied in all possible non-crystalline phases by experiments, theoretical approaches and computer simulations. The short-range structural order is rather straightforward as only one type of bonding appears (Se–Se) with bond length $r \approx 2.3$ Å, mean coordination number $<n> \approx 2.0\pm0.1$, bond angle of $105^{o}$ and dihedral angle in the range $70^{o} - 100^{o}$. The dihedral angle is the only among the aforementioned structural parameters that differentiates glassy and crystalline t-Se ($102^{o}$). Open issues are therefore related to the medium-range structural order and the clarification of the ring/chain equilibrium, interchain interactions, and so on. Representative studies include inelastic neutron scattering [52,53,54], inelastic X-ray scattering [55], X-ray absorption fine structure [56,57], X-ray and neutron diffraction [58,59], viscosimetry [60,61], ultraviolet photoemission spectroscopy [62], nuclear magnetic resonance [63], reverse Monte Carlo [64,65] and various classical molecular dynamics and *ab initio* molecular dynamics simulations [66,67,68,69,70,71]. Vibrational spectroscopies (Raman and IR) have also provided a great deal of information about structural correlations at various length scales [72,73,74,75,76,77,78,79,80].

The identification of the Raman modes of amorphous selenium has been a long-standing and perplexing issue. Early studies have misinterpreted the origin of the peaks in the high frequency are, $200-300$ cm$^{-1}$. For years it was erroneously considered that the two peaks at ~235 and ~250 cm$^{-1}$ have been the signatures of vibrations of long Se$_n$ chains and Se$_8$ rings, respectively [72,75]. Indeed, the first attempt to assign the Raman bands of amorphous Se was reported by Lucovsky *et al*. [72]. Comparing the spectrum of the amorphous film and the $\alpha$-monoclinic crystal, which is composed of Se$_8$ rings, Fig. 10(a), they concluded that both rings and chains are present in amorphous Se. However, it was later appreciated that the A$_1$ symmetric vibrations of both rings and chains are found near the 250 cm$^{-1}$ frequency [81]. This conclusion was considered to arise from the fact that in non-crystalline Se rings and chain fragments act essentially as isolated species having much weaker inter-molecular interactions in comparison to the crystal.

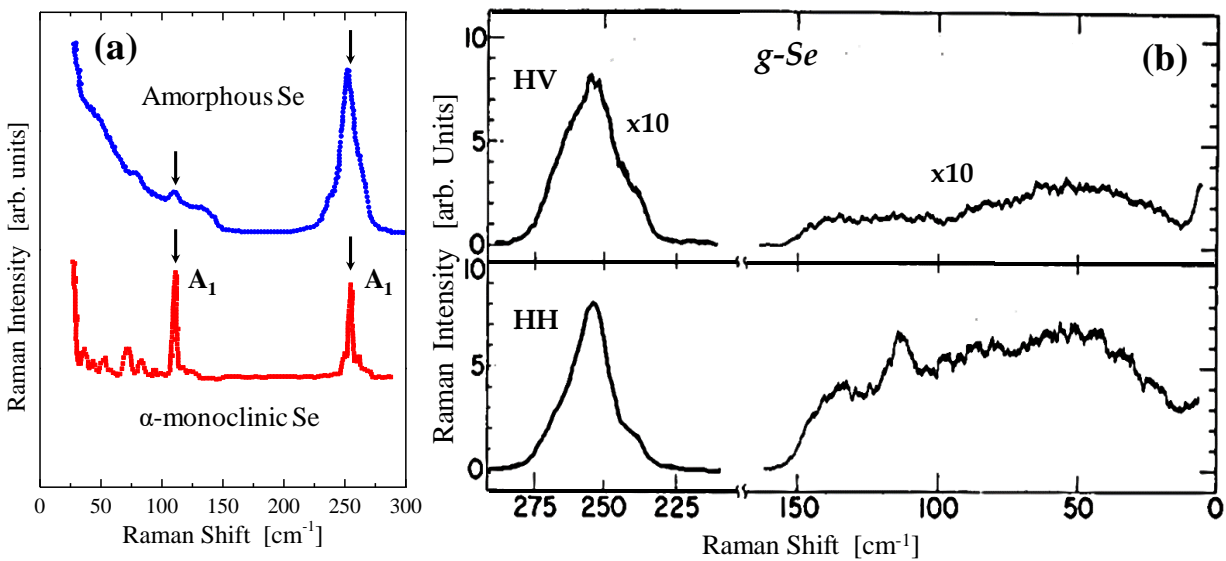

**Fig. 10:** **(a)** Raman spectra of amorphous and α-monoclinic Selenium. Adapted from Ref. [75]. **(b)** Polarized (HH) and depolarized (HV) Raman spectra of bulk glassy Se recorded at 9.8 K using the 799.3 nm as the excitation source. Reprinted from Ref. [74]. *Copyright © 1976 Published by Elsevier Ltd.*

Gorman and Solin [74] recorded for the first time the polarized and depolarized Raman spectra of glassy Se at low temperature (9.8 K) with excitation energy (1.55 eV) much lower than that of the optical bandgap (~2.0 eV), as shown in Fig. 10(b). The polarization analysis shows that the main Raman band at 250 cm$^{-1}$ is strongly polarized with a depolarization ratio of ~0.1. The authors also criticized Lucovsky's assignment of the 250 cm$^{-1}$ as a ring mode based on the mere coincidence of the band of monoclinic Se. Misawa and Suzuki [82] and Lucovsky [75] proposed independently that the structure of non-crystalline Se contains – apart from isolated Se$_8$ rings and Se$_n$ chains – contains molecular fragments, which are incorporated into the chains having ringlike conformations.

Carroll and Lannin [77] showed that the structure of amorphous evaporated Se is almost identical to that of the bulk glassy Se. Their Raman spectra are shown in Fig. 11 where the 112 cm$^{-1}$ mode, representative of Se$_8$ rings is present in both spectra with similar intensity. This comparison casts solid evidence against succeeding studies, even recent ones, supporting that the structure of amorphous Se is dominated by Se$_8$ rings. Carroll and Lannin provided evidence to support that the 112 cm$^{-1}$ band arises from Se$_8$ ring, hence, criticizing the Lucovsky's [75] suggestion who considered that this band was related to ring-like fragments found into chains.

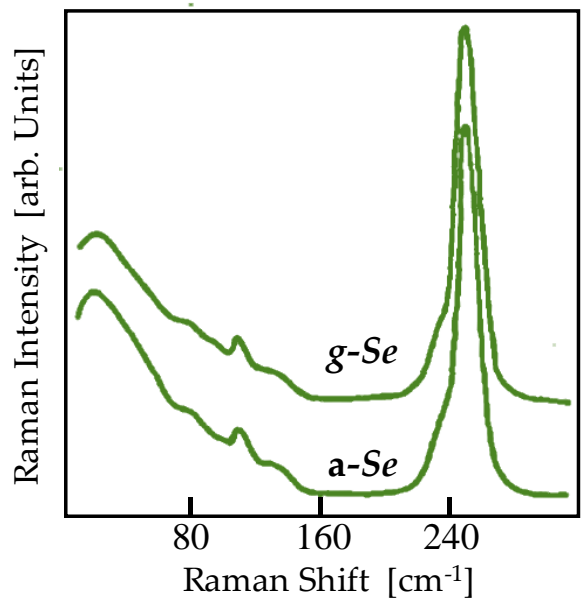

**Fig. 11:** Raman spectra of bulk glassy and evaporated amorphous Se. Adapted from Ref. [77]. *Copyright © 1981 Published by Elsevier Ltd.*

The initially erroneous assignment of the Raman bands [72] has proliferated through the succeeding works adding confusion and leading to conclusions that are not essentially supported by the experimental data. Carini *et al.* [76], based on the assignment of the 235 and 250 cm$^{-1}$ bands to chains and rings, respectively, attempted to estimate the temperature-dependent relative fraction of these species using the ratio of the 235 cm$^{-1}$ band over the total intensity including both bands. The rapid increase of this intensity ratio above $T_g$ was incorrectly attributed to the increased Se$_8$ concentration. However, obviously this change simply signifies the onset of crystallization. Further, an alleged resonant Raman scattering study of glassy Se identified new Raman bands at 31, 62 and 93 cm$^{-1}$ assigning them to ring structures [83]. This led the authors to propose a method to quantify the ring-to-chain equilibrium. As Griffiths showed [84] the above triplet of bands are artifacts arising from intracavity modes of the dye laser.

A comparison with the true VDoS shows that the Raman spectrum reflects rather accurately the entire density of states including torsional, bending and stretching modes, as shown in Fig. 12(a). The Raman bands at high energy (200-300 cm$^{-1}$) are sharper than the VDoS bands, implying that the Raman coupling coefficient of certain stretching modes dominates. A detailed analysis of the reduced polarized Raman spectra, in analogy to the analysis of liquid sulfur, revealed that three bands are hidden below the spectral envelope spanning the wavenumber range 220−280 cm⁻¹ [79], see Fig. 12(b). The bands correspond to t-Se chains at ~234 cm⁻¹, Seₙ chains in disordered configurations at ~250 cm⁻¹ and Se₈ rings at ~260 cm⁻¹. This result is in agreement with the Raman results of elemental Se confined in nanochannels of AlPO₄-5 single crystals [85]. It is worth mentioning that despite the clarification of the origin of the Raman bands several publications, even recent ones, continue to misinterpret the 250 cm⁻¹ with ring or monoclinic content. A typical example is found in a high-pressure Raman study [86] where purely crystalline Se with a Raman band at 234 cm$^{-1}$ was considered as amorphous. The wrong interpretation cast serious doubts on the alleged phase transitions observed under temperature and pressure variation.

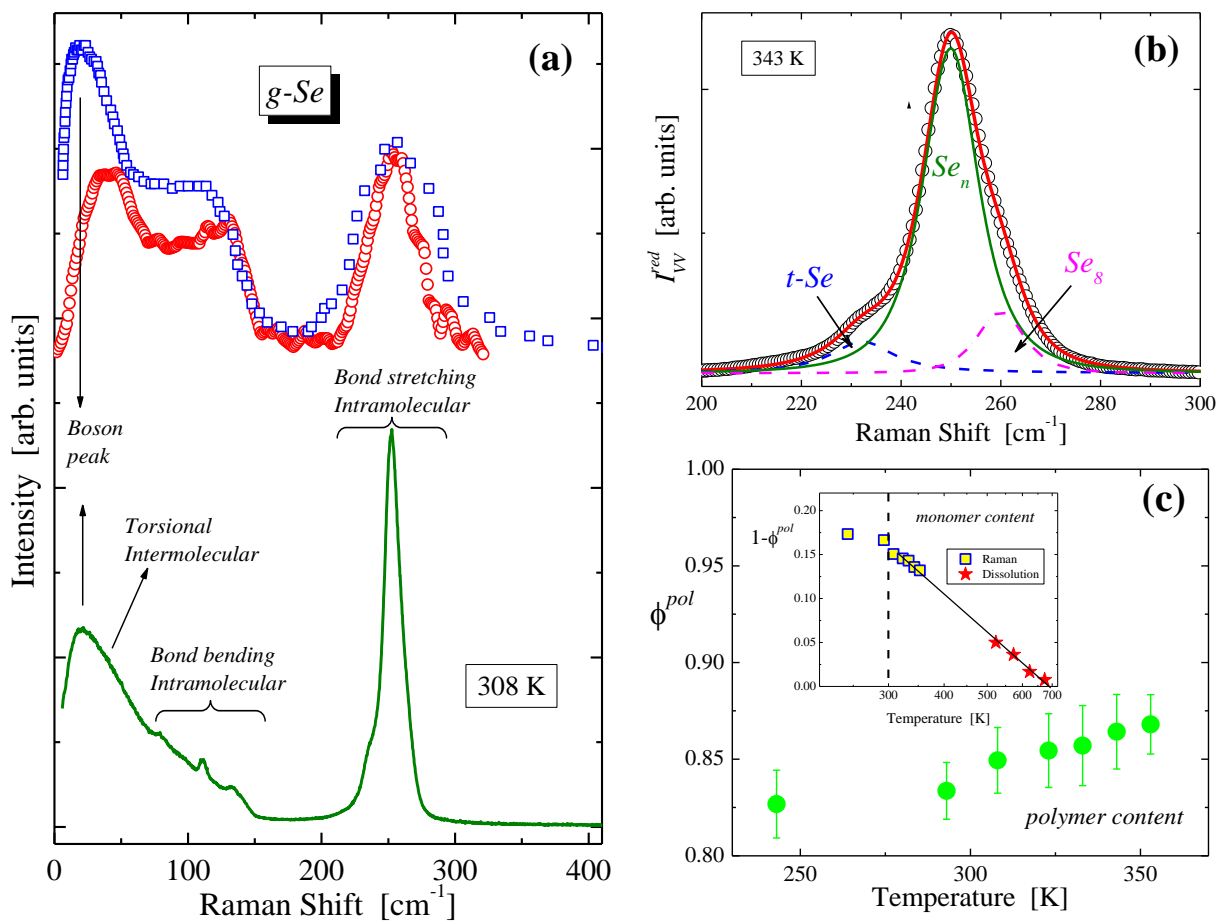

**Figure 12:** **(a)** Raman spectrum of g-Se near $T_g$ (solid line) recorded at sub-bandgap conditions (752.4 nm). Vibrational density of states taken from Ref. [53] (open squares; Bose scaled) and Ref. [54] (open circles). **(b)** Fitting analysis of the bond stretching frequencies. **(c)** Temperature dependence of the weight fraction of *Se* atoms participating in polymeric chains. The inset shows the monomer content (atoms in Se₈ rings) in a semilogarithmic plot. Low temperature: Raman data (squares); high temperature: quench-and-dissolution data [88]. "Reprinted from [S.N. Yannopoulos, K. S. Andrikopoulos, J. Chem. Phys. 121, 4747–4758 (2004)], with the permission of AIP Publishing."

The analysis of the vibrational modes, shown in Fig. 12(b), enabled the calculation of the temperature dependence of the polymer fraction of glassy and supercooled Se. The latter was estimated by the peak intensity ratio as $\phi^{pol} = I^{Se_n}/(I^{Se_n} + I^{Se_8})$. The temperature dependence of $\phi^{pol}$ is shown in Fig. 12(c). The $\phi^{pol}(T)$ curve exhibits slope change near $T_g$. This may be attributed to the unfreezing of diffusional motion, which renders the ring-to-chain transformation more facile. This change has immediate consequences also for the modifications observed at the medium-range structural order upon temperature variation [78]. The analysis of the composite bond-stretching band demonstrated that the presence of the 112 cm$^{-1}$ in the intermediate or bond-bending frequency range, is not the only direct proof of the existence of Se$_8$ rings in the glass structure.

The magnitude of the monomer content in Se obtained by analyzing Raman spectra is significantly lower than that resulting by early quench-and-dissolution methods [87]. Applying the theory they originally developed for the $\lambda$−transition of sulfur, Eisenberg and Tobolsky [16] estimated the polymerization transition temperature of Se at $T_\lambda^{Se} \approx 356$ K, which is ~138 K below the melting point. However, this calculation has been debated [75] in view of the large uncertainties accompanying Briegleb's data [87] about the monomer content of Se. The structure of glassy selenium and the ageing it experiences at ambient conditions was recently studied by Boolchand and coworkers with calorimetric methods and Raman scattering [80]. Analyzing Raman spectra in the spirit illustrated in Fig. 12(b) it was observed that both the t-Se and Se$_8$ components increase in intensity upon ageing time (several months) at the expense of the intensity of the main Se$_n$ band.

Misawa and Suzuki were the first to suggest that the structure of glassy/amorphous Se is not simply described as a mixture of isolated rings and chains but contains, in addition, fragments characterized by both ringlike and chainlike conformations [82]. A key ingredient of their model is the presence of threefold–coordinated Se atoms, which may be the cause of several remarkable phenomena that g-Se exhibits. Popov [88] performed more accurate experimental measurements of the monomer content in Se. The data are plotted in the inset of Fig. 12(c) alongside with the Raman results. The two data sets seem to conform remarkably. Another estimation of the monomer content using Raman spectra was undertaken by comparing the intensities of the peaks located at 112 cm$^{-1}$ (rings) and 250 cm$^{-1}$ (chains) [77]. This estimation is risky and probably of rather limited reliability as the authors employed the comparison of the bond-bending mode of rings and the bond-stretching modes of chains; whose Raman coupling coefficients differ substantially.

A wealth of information on the medium-range structural order can be gained by studying the bond-bending spectral range, 70−150 cm$^{-1}$ [71,78]. Detailed molecular dynamics simulations have suggested a correlation among the relative intensity of peaks in that spectral area and the parameter $f_{RRR}$ which was related to the steric hindrance effect. $f_{RRR}$ was selected to control the chain conformations and hence the medium-range structural order. Temperature is the experimental parameter that can take the role of $f_{RRR}$, as steric hindrance can be dictated by temperature itself. On the experimental side, the analysis is based on the sharp bands at 112 and 138 cm$^{-1}$, Fig. 13(a), which arise from bond-bending vibrations of ringlike and helixlike species, respectively. The intensity ratio of the bands at 112 and 138 cm$^{-1}$, $I^{112}/I^{138}$, exhibits moderate increase upon heating the glass. The rate of this increase changes considerably at $T > T_g$. The experimental results are qualitatively similar with the results obtained by molecular dynamics simulations [71], see Fig. 13(b). This shows that the population of ringlike arrangements increases with increasing temperature. The experimentally obtained activation energy for this transformation amounts to $\Delta E \approx 0.2$ eV [78], which is in striking agreement with the value resulting from simulations, $\Delta E \approx 0.1$ eV [71].

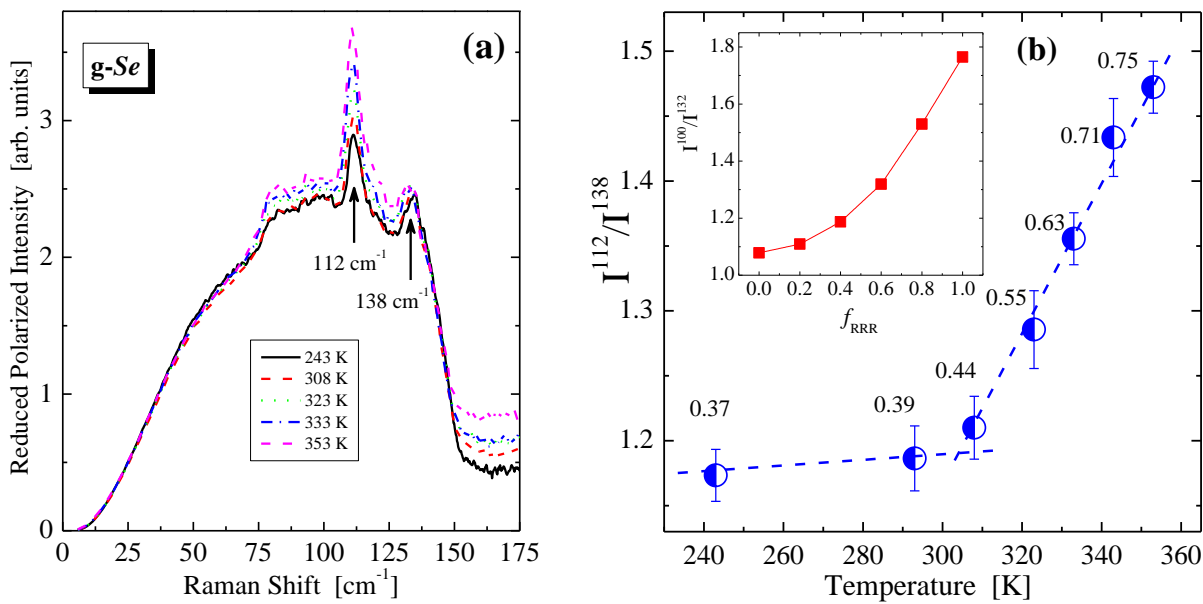

**Figure 13: (a)** Low frequency reduced polarized Raman spectra of g-Se at various temperatures. **(b)** Temperature dependence of the intensity ratio of the 112 and 138 cm$^{-1}$ Raman bands. Dashed lines are guides to the eye to show the slope change when crossing $T_g$. The inset shows the dependence of the corresponding ratio obtained by theoretical analysis taken from Ref. [71]. "Reprinted (figures 2, 3) with permission from [S.N. Yannopoulos, K. S. Andrikopoulos, Phys. Rev. B 69, 144206 ] Copyright (2004) by the American Physical Society."

Exploration of vibrational properties of Se with surface-sensitive structural probes corroborated that surface bonding may differ from the bulk. This was a well-conceived knowledge for crystals, while no analogous evidence was provided for amorphous solids. Amorphous Se films prepared by pulsed-laser deposition were investigated using grazing incidence inelastic X-ray scattering and computer simulations [89]. An increased VDoS was found in the energy region between 17 and 28 meV (or $140 - 225$ cm$^{-1}$), which is the gap that separates acoustic and optic modes in the bulk vibrational density of states, see Fig. 14(a). These additional states arise from the depletion of vibrational modes at the other parts of the spectrum (mainly torsional). *Ab initio* and semi-empirical molecular orbital theoretical calculations shed light on the origin of these states by assigning them to hypervalent defects, i.e. pairs of 3-fold and 4-fold coordinated Se atoms. This branching (inter-chain bonding) of the 1-D Se chains leads to the formation of a few nm surface layer with enhanced rigidity.

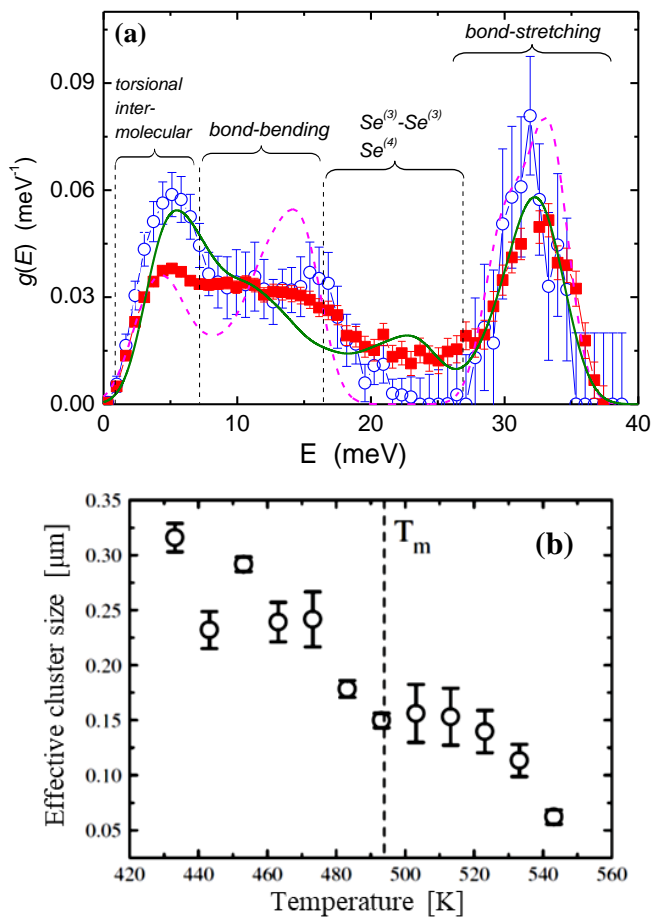

**Figure 14: (a)** Vibrational density of states, $g(E)$, of a-Se measured in bulk (open circles) and surface-sensitive geometry (solid squares). The solid and dashed lines represent the simulated $g(E)$ of the surface of a Se$_{96}$ cluster and a bulk Se$_{96}$ chain, respectively. Reprinted from Ref. [89]. *Copyright © 2011, Springer Nature.* **(b)** Temperature dependence of the "hydrodynamic: radius of Se cluster in the supercooled and molten state. Reprinted from Ref. [90]. *Copyright © 2009 Elsevier B.V. All rights reserved.*

Using photon correlation spectroscopy operating at near-IR wavelengths, Cazzato *et al.* [90] were able to explore the dynamics of Se in the supercooled and molten state. The liquid was found to exhibit a single relaxation process which was assigned to the self-diffusion dynamics of Se clusters. The effective "hydrodynamic" radii, Fig. 14(b), of the clusters exhibit strong temperature dependence decreasing from 350 nm at 430 K to 50 nm at 430 K.

Closing this section, it worth mentioning that a large body of information has been obtained by exploring various molecules or clusters of Se confined into various matrices, such as zeolites [91,92,93] and more recently in carbon-based nanostructures [94,95,96]. Isolation into zeolites may infer information related to the structure and conformation of isolated structures (rings, chains) given that the interaction with the pore walls are negligible. Confinement into carbon-based structures is mainly investigated towards improving storage capacity of batteries and related energy storage devices.

## 7. Tellurium

Much less attention has been paid to elemental Te in relation to its chalcogen counterparts, S and Se. The considerably lower homonuclear bond energy in Te in relation to S and Se, might explain the high lability of Te to external stimuli, such as its proclivity to oxidation and its low threshold ablation under laser light leading to growth of anisotropic nanostructures [97,98]. Te has strong tendency in forming chains (hexagonal grey Te) instead or homocyclic closed ring molecules. Secondary interactions offer interchain links which are longer than covalent (intrachain) bonds, but shorter than the sum of van der Waals radii.

### 7.1 Amorphous Tellurium

Elemental Te cannot be obtained in the bulk glassy state even if high quenching rates are applied [5,6,99]. There are, surprisingly, a few reports on rapid solidification of the melt towards the formation of the bulk glass [100,101]. The results of these reports are highly questionable since the polycrystalline form is most probably formed by melt quenching, instead of the glass. Amorphous films can be prepared by physical vapor deposition techniques given that the substrate is cooled at low enough temperatures to avoid spontaneous crystallization. Amorphous Te was recently obtained by irradiation of the crystal with strong femtosecond pulses arising from thermal melting [102]. The amorphous phase was stable only at temperatures lower than 200 K, while fast crystallization takes place at room temperature. The process follows Arrhenius dependence with an activation energy of ~0.82 eV.

Amorphous Te is essentially unsuitable for practical applications due to its limited thermal stability; it crystallizes at ~10 °C [103]. Based on reflectivity studies it has been found that the energy gap for the absorption edge almost doubles going from the crystal (0.4 eV) to the amorphous phase (0.8 eV) [104]. This change results in a large contrast of the reflectivity between the two phases. The large change of the optical constants of a-Te in comparison to t-Te has been assigned to the weakening of interchain interactions in the amorphous solid [103]. This endows a-Te a more molecular-like character than that of t-Te. Ichikawa was the first who performed a diffraction study to elucidate the structure of a-Te [105]. The mean coordination number of the nearest neighbors was estimated to be ~1.7, which led to the conclusion that chains in a-Te are short, composed of nearly 10 atoms per chain. An NMR study was not very conclusive as the putative glass obtained by quenching of Te melt was found to contain both amorphous and polycrystalline parts [100]. More recent EXAFS studies revealed that the intrachain coordination number of a-Te is close to 2.0, hence, discarding the idea of very short length chains [106].

Amorphous Te has been the subject of structure-probing techniques such as photoemission spectroscopy [107,108,109], Mössbauer [110], and computer simulations [111,112]. Akola and Jones presented the most detailed study based on density functional and molecular dynamics simulations [112] exploring the amorphous and the liquid state, as well as the stability of various $Te_n$ ($n = 2-16$) ring molecules. Notably, there is currently only one Raman study of amorphous Te conducted more than 45 years ago by Brodksy *et al*. [113]. The comparison between the Raman spectra of t-Te and a-Te are shown in Fig. 15(a), while the reduced spectrum of the amorphous sample is shown in Fig. 15(b). The symmetries and selection rules for the space group of t-Te dictate that three out of the four vibrational modes are Raman active. The irreducible representation for t-Te is composed of four optic modes, two non-degenerate and two doubly degenerate, and can be written as $A_1 + A_2 + 2E$ [114]. The atomic motion of the optic modes is shown in Fig. 15(c). $A_2$ represents the rigid chain rotation (IR-active) and is affected solely by interchain interactions.

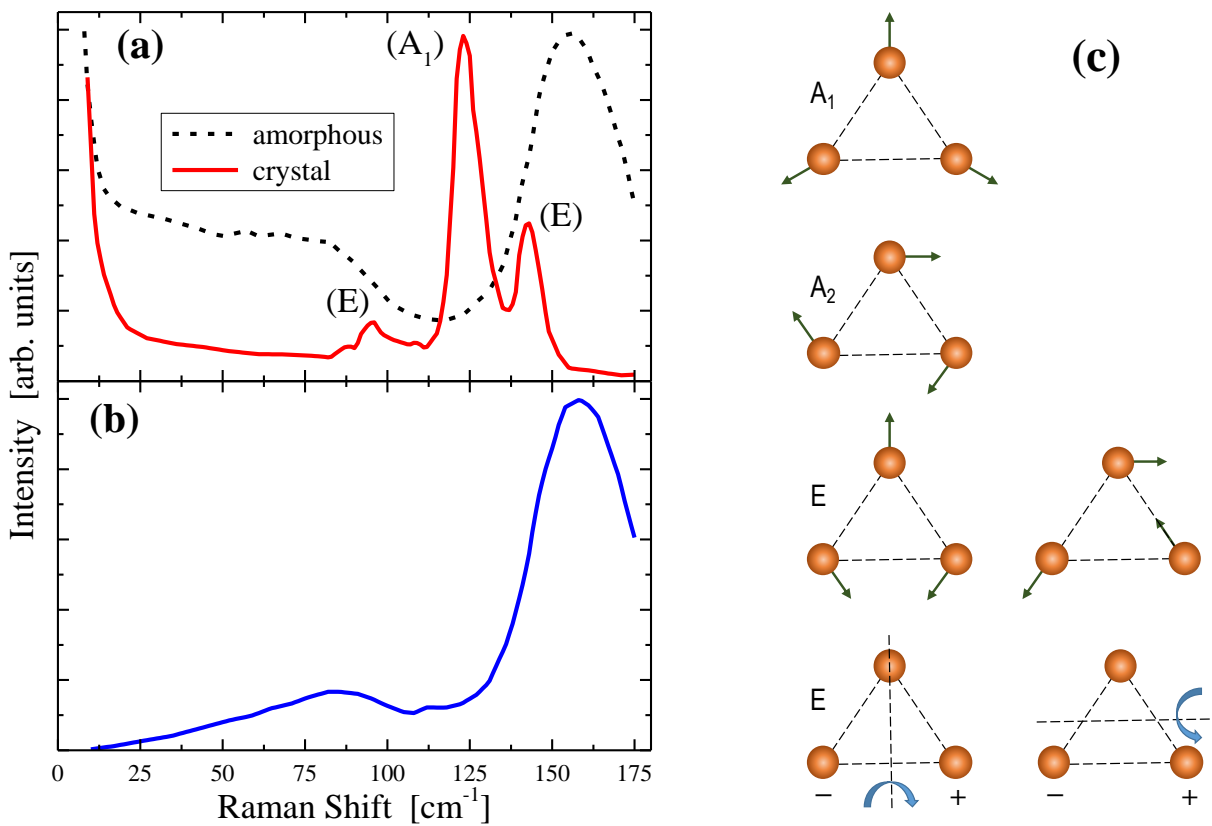

**Figure 15:** **(a)** Raman spectra of polycrystalline (solid line) and amorphous Te (dashed line). Spectra have been measured at 77 K. Excitation wavelength: 514.5 nm, spectral resolution 8 cm$^{-1}$. **(b)** Reduced Raman spectrum of amorphous Te. Reprinted from Ref. [113]. *Copyright © 1972 WILEY-VCH Verlag GmbH & Co. KGaA.* **(c)** Schematic representation of the atomic motion of the optic modes of the trigonal chain. $A_1$: symmetric breathing mode. $A_2$: rigid chain rotation. E (upper): asymmetric stretching. E (bottom): intrachain rotation.

In analogy to the case of elemental Se, the Raman spectrum of a-Te was interpreted as providing evidence for an appreciable weakening of interchain forces when going from the crystal to the disordered phase. As a result, the intrachain bond strengthens and the corresponding mode $A_1$ shifts from 122 to 157 cm$^{-1}$. The interchain bond length in t-Te is only 23% longer than the intrachain bond length. While in Se the ratio of the $A_1$ mode between the amorphous and the crystal is $\nu(A_1^{am})/\nu(A_1^{cr}) \approx 1.07$, it exhibits much larger value for Te, i.e. $\nu(A_1^{am})/\nu(A_1^{cr}) \approx 1.29$.

### 7.2 Liquid Tellurium

Due to the complexity to produce and maintain amorphous Te at sufficient quantities for diffraction or other structure-probing experiments, the great preponderance of structural information has been offered by studying liquid Te. Deeply supercooled liquid Te exhibits notable changes in various physical properties such as thermal expansion, compressibility, and specific heat at $T^*\sim353\pm3$ ℃, namely, almost 100 ℃ below the melting point [115]. Taking into account previous diffraction studies, a model has been proposed for molten Te where the high temperature liquid (900 ℃) is a 3-fold coordinated network structure, while near the melting point an increasing number of Te atoms attain 2-fold bonding [116]. Therefore, liquid Te was envisaged as a network of covalent bonds coexisting with a metallic-like electron gas. Additional subsequent diffraction studies perplexed this model suggesting that the shape of the radial distribution function is not proper to unequivocally extract the coordination number from such data [117]. The unexpected changes in physical properties mentioned above were thus considered to arise form a structural phase transition where Te atoms changes coordination number from 3 at $T > T^*$ to 2 at $T < T^*$. In addition, liquid Te near its melting point exhibits a density maximum; a well-known anomaly also observed in water. Combined density functional/molecular dynamics simulations have concluded that both 2-fold and 3-fold coordinated Te atoms are present in the liquid and temperature dependent alterations of chain lengths, ring distributions, and cavity volumes are responsible for the density anomaly [118].

Raman spectra recorded at various temperatures in the molten state up to 530 ℃ showed minor spectral changes in comparison to the crystal [119]. Representative Raman spectra are shown in Fig. 16(a). The main changes taking place upon increasing temperature are: (i) a broadening of the Raman peaks of the crystal and (ii) a progressive red-shift of the band energies. The main $A_1$ symmetry band shifts from 122 cm$^{-1}$ in the crystal to 116 cm$^{-1}$ in the liquid at 530 ℃. The very close resemblance of the spectra from ambient temperature to the melt led the authors to conclude that no drastic structural changes exist between the crystal and the melt; hence suggesting a 2-fold chain model for the melt. Additional Raman studies including polarization analysis were reported by Magana and Lannin [120]; the polarized (HH) and depolarized (HV) spectra of the melt are shown in Fig. 16(b). The $A_1$ band at ~112 cm$^{-1}$ is strongly polarized while the other two modes are depolarized. Although the Raman spectra of molten Te provided by these two investigations are consistent, a different interpretation was provided by Magana and Lannin who considered that molten Te is described as a continuous random network dominated by the 3-fold coordination of Te atoms. The authors developed arguments to ground their conclusion of a 3-fold coordinated structure into the melt, based on comparison with the Raman spectra of supercooled Se. In addition, comparison between vibrational frequencies of *l*-Te with those of the pyramidal units in a-As and a-P, were considered as further support to the 3-fold Te coordination [120]. However, the interpretation of Raman spectra based on the above mentioned reasoning cannot be considered as solid evidence for supporting the model of 3-fold coordinated Te in the molten state.

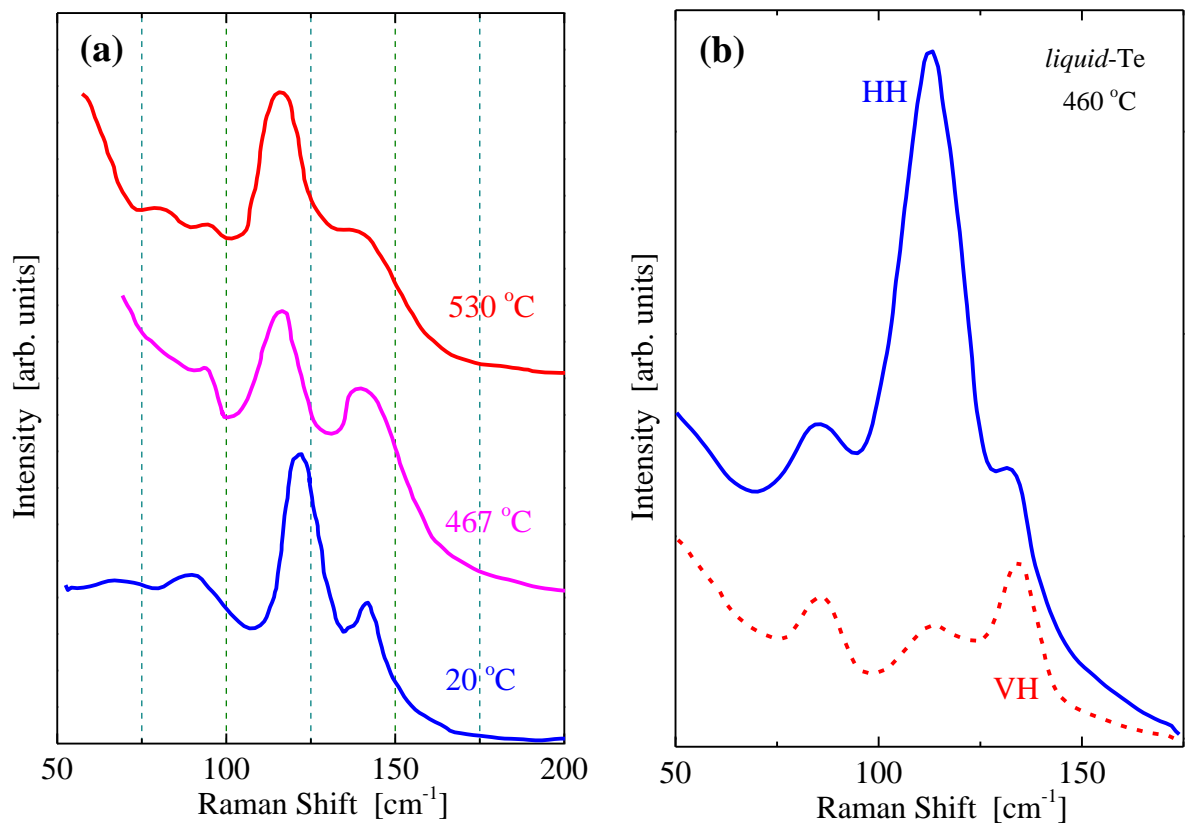

**Figure 16** **(a)** Raman spectra of crystalline and molten Te. Curves were digitized from Ref. [119]. © Springer-Verlag Berlin Heidelberg New York, 1979. **(b)** Polarized (solid line) and depolarized (dashed line) Raman spectra of *l*-Te at 460 °C. The intensities of the VH spectra have been normalized to match the depolarization ratio values provided in the original article. Curves were digitized from Ref. [120]. https://doi.org/10.1103/PhysRevLett.51.2398.

It still remains a puzzling issue to comprehend why the Raman spectrum of *l*-Te shares several common features with that of crystalline *t*-Te, being also quite dissimilar with the spectrum of amorphous a-Te. In conclusion, *l*-Te is a counterintuitive example to the notion that the amorphous and liquid states show very similar Raman spectra. As *l*-Te is highly absorbing any excitation source that can be used in Raman spectroscopy, other phonon-probing techniques can be employed.

To resolve the above issue, more solid evidence about the local atomic arrangement in *l*-Te has been provided by inelastic neutron scattering experiments [121] which have measured the VDoS. As stated in Section 3, Raman scattering in non-crystalline media (amorphous solids and melts) is propositional to the VDoS modulated by the coupling coefficient; hence, care should be exercised in comparing band intensities. The VDoS reported by Endo *et al*. [121] was measured at 400 °C for the crystal and the supercooled liquid and at 467 °C for the melt, see Fig. 17. The VDoS of t-Te shows three main bands at 5, 10 and 16 meV (1 meV $\hat{=}$ 8 cm$^{-1}$) assigned to torsional, bond-bending and bond-stretching modes, respectively. The presence of these three zones in the VDoS of supercooled Te at 400 °C supports the preservation of the chain structure in the low-temperature liquid. The three zones are also evident at 467 °C despite the broadening of the VDoS features. The torsional modes are almost unchanged, while the bond-stretching density of states shifts to lower energies (13 meV), which was assigned to longer bonds in accordance with previous diffraction studies interpreted in terms of short and long bonds.

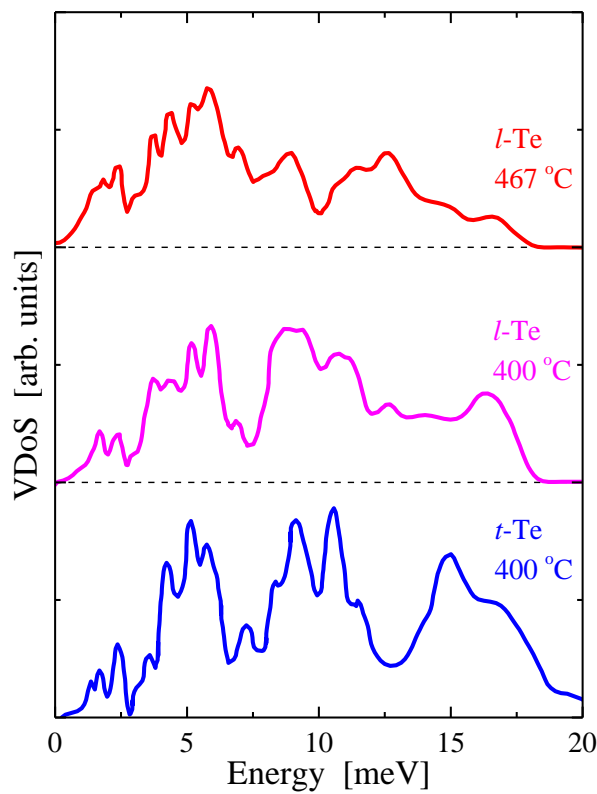

**Figure 17:** Vibrational density of states of t-Te and l-Te at various temperatures. Curves were digitized from Ref. [121]. *Articles under CC BY 4.0 © [1994] The Author(s)*. Originally published in J. Phys. Soc. Jpn. 63, (1994), Quasielastic and Inelastic Neutron Scatterings of Liquid Tellurium, 10.1143/JPSJ.63.3200.

### 7.3 Tellurium nanoparticles

Confinement of matter to the nanoscale brings about changes in structure and bonding mimicking disorder when the surface-to-volume atomic ratio becomes important. In this context, nanomaterials are typically considered as being situated between ordered (crystalline) and disordered solids. The difficulty in studying amorphous Te, owing to its very low thermal stability and the failure to produce the bulk glass, has led to the exploration of various nanostructures where incipient disorder effects accompany size shrinkage. The chain-like structure of t-Te is the decisive factor justifying its high propensity to develop anisotropic 1-D structures, either in solution-based growth or by ablation/deposition methods [97]. However, selecting proper conditions, growth of spherical Te nanoparticles can also take place [122].

In an effort to investigate the influence of nanoscale on the conformation of Te chains, Ikemoto and Miyanaga used EXAFS to study alternating layers of ultrathin Te layers (fragmented into various sizes of nanoparticles) separated by layers of NaCl [123]. A marked outcome of this study is that the intrachain covalent bond strengthens as it shrinks by 0.042 Å in comparison to the bond length in bulk t-Te. This arises from the substantial reduction of the average interchain coordination number, which halves, from ~4.2 in the bulk to ~2.2 at the nanoscale. Preparing thin films of various thicknesses, they suggested correlations of the interchain first nearest neighbor coordination number ($N_{inter}$) with the first nearest neighbor intrachain distance $r_{intra}$, and the force constant, as shown in Fig. 18(a). Further, exploring the variation of the intra-to-inter-chain interactions as a function of the mean nanoparticle size, it was suggested that nanoparticles smaller than ~60 nm are mixtures of chains in crystalline and amorphous conformations, while for large particles the crystalline component prevails [124].

In a recent study, growth of Te nanoparticles was achieved by cw laser irradiation of bulk t-Te crystals. Contrary to the growth of 1-D morphologies (including nanosized tubes, rods and wires) favored by the stability of chain in the crystal, the gas-assisted transfer of the laser-produced vapors onto another substrate results in the formation of spherical nanoparticles [122]. Raman spectra of such nanoparticles with an average diameter of ~100 nm are shown in Fig. 18(b). The comparison between the Raman spectra of bulk t-Te and the nanospheres reveal significant line broadening of the bands for the latter. In addition, a weak broad background also develops in the case of nanospheres. These spectral changes have been attributed to the shortening of *t*-Te chains and the disordering effects caused by the strain that develops within the spherical crystallites [122].

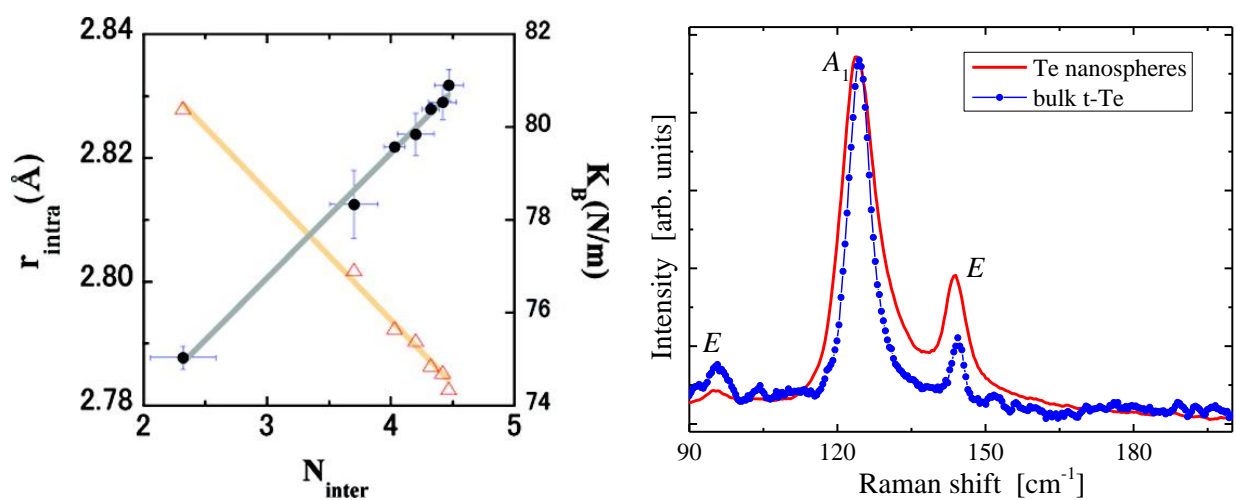

**Figure 18: (a)** Correlations of the interchain first nearest neighbor coordination number ($N_{inter}$) with the first nearest neighbor intrachain distance $r_{intra}$, (solid circles) and the force constant $K_B$ open triangles.

## 8. Binary chalcogen glasses

### 8.1 Sulfur−Selenium binary glasses

Binary S-Se compounds have mainly been studied in their crystalline state by various techniques including Raman spectroscopy [50, 125, 126]. To assign Raman bands appearing in the spectra of mixtures, Ward proposed an empirical model [50] taking into consideration scaling arguments based on the masses of the ring and force constants dependent on the relative S/Se ratio in each mixed ring. Based on Raman spectra [125] and considering that the dissociation energies of chalcogen bonds follow the order S−S > Se−S > Se−Se it was suggested that Se−Se bonds are preferable in cyclic rings at fractions larger than those dictated by simple statistical mixing. Indeed, Se−Se bonds in $Se_nS_{8-n}$ rings can be found in S-rich mixtures since an $SeS_7$ ring can be attacked more easily by another Se atoms in relation to an $S_8$ ring. On the Se-rich limit of the Sulfur-Selenium binary mixture, S atoms prefer a more statistical mixing and S-S bonds are less preferable for the same reasons. NMR studies of Sulfur-Selenium melts showed the predominance of $SeS_7$ and $Se_2S_6$ over a composition range up to 45 at. % Se, while other cyclic molecules exist at much lower concentrations [127]. The polymeric content was interpreted as a statistical random distribution of selenium and sulfur atoms in chains. The increase of Se content in the binary mixture resulted in an increase of the polymeric content at the expense of the fraction of rings [127].

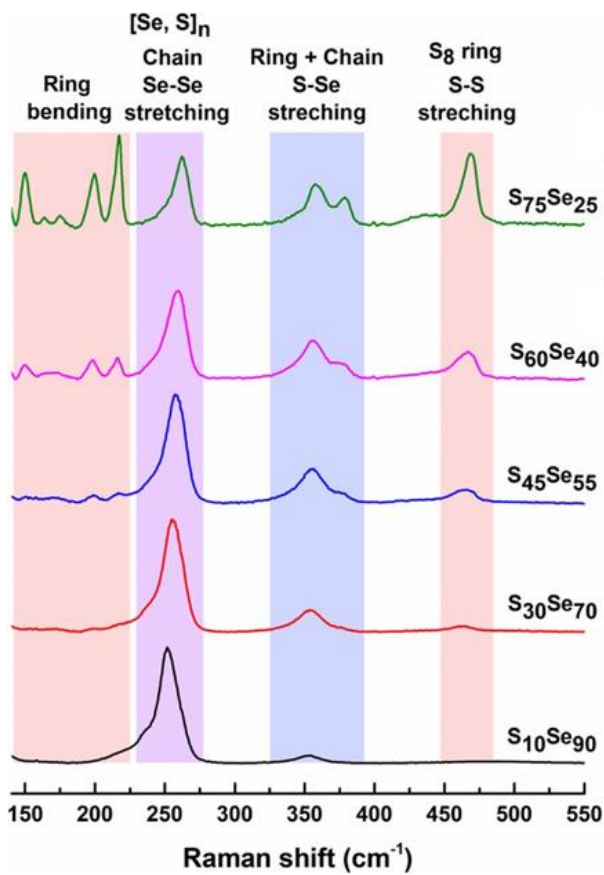

**Fig. 19**: Raman spectra of binary $S_xSe_{100-x}$ glasses at various compositions. "Adapted with permission from (Structure and Chemical Order in S–Se Binary Glasses, B. Yuan, et al., J. Phys. Chem. B (2018) **122**(50), 12219-12226)." *Copyright © (2018), American Chemical Society.*

A series of binary bulk glasses over the composition range $S_xSe_{100-x}$ ($0 \leq x \leq 90$) was recently prepared and studied by Raman and NMR spectroscopies [128]. Mixed eight-membered rings and mixed polymeric chains are the main constituents of the glass structure. Raman spectra of binary glasses over a broad composition range are shown in Fig. 19(a). New bands appearing at 163, 175, and 202 cm$^{-1}$ were assigned to bending vibrations of $Se_3S_5$, $Se_2S_6$, and $SeS_7$ rings, respectively, based on Ward's model mentioned above. The Raman spectra revealed that ring molecules with more than one Se atoms can be found even at the mixtures with higher S content namely $S_{75}Se_{25}$. The progressive increase of S into Se chains has resulted in a gradual shift of the main vibration mode of Se−Se bonds in disordered chains from ~250 cm$^{-1}$ in pure Se glass to ~260 cm$^{-1}$ in the $S_{75}Se_{25}$ glass.

### 8.2 Selenium−Tellurium binary glasses and films

Owing to the inability of elemental Te to be prepared as a bulk glass, the $Se_xTe_{100-x}$ binary system can essentially be vitrified only in the Se-rich phase, i.e. x ≥ 80 using typical laboratory quenching rates [5]. In a recent detailed study of the reversible amorphous-to-crystalline phase change using ultrafast differential scanning calorimetry it was reported that purely non-crystalline products can be obtained up to the $Se_{60}Te_{40}$ composition, at a quenching rate of 4000 K s$^{-1}$ [129]. For this high quenching rate, the binary compound becomes progressively crystalline in the range 40 < x < 60, being totally crystalline at this upper limit. The critical cooling rate to prevent crystallization changes exponentially with the Te content ranging from 10 K s$^{-1}$ at x∼80 at.% of Te to 6000 K s$^{-1}$ at x∼40 at.% of Te [129].

While the Raman studies dealing with the Sulfur-Selenium binary glassy system are rather scarce, more attention has been paid to non-crystalline (films and glasses) Selenium-Tellurium compounds despite the inferior glass-forming ability of the latter. Geick *et al*. [130] analyzed the Raman spectra of the $Se_xTe_{100-x}$ binary crystals straddling the whole composition range from Se to Te. The experimentally observed Raman bands were analyzed in relation to the random element isodisplacement model.

Ward [131] explored Se-rich $Se_xTe_{100-x}$ (x = 5, 10) glasses by Raman spectroscopy. Representative Raman spectra are shown in Fig. 20(a). The Raman mode emerging at ∼216 cm$^{-1}$ was erroneously assigned to mixed $Se_2Te_6$ rings. This assignment was biased by the misleading notion that time about the origin of the 250 cm$^{-1}$ band of glassy Se, which was attributed to $Se_8$ ring molecules. This point was confidently clarified in later Raman studies [79] as discussed into detail in Section 6.1. Polarized and depolarized Raman spectra of Te-rich Selenium-Tellurium binary liquids were recorded near the melting point [132]. The data were interpreted in the context that liquid *l*-Te is 3-fold coordinated. The analysis was based on analogies between the Raman spectrum of *l*-Te, which appeared to be qualitatively similar – as regards peak frequencies and polarization of peaks – to that of amorphous As and P. While the Raman spectra of *l*-$Se_{20}Te_{80}$ exhibit strong similarity (apart for moderate blue-shits) with *l*-Te, abrupt changes were observed in the melt structure for Te content around 30 at.%. These significant spectral changes were assigned to a structural change from 3-fold to 2-fold coordination of Te atoms given that other physical properties also change near this composition. A small fraction of 2-fold coordinated Te atoms for x ≤ 20 was not precluded.

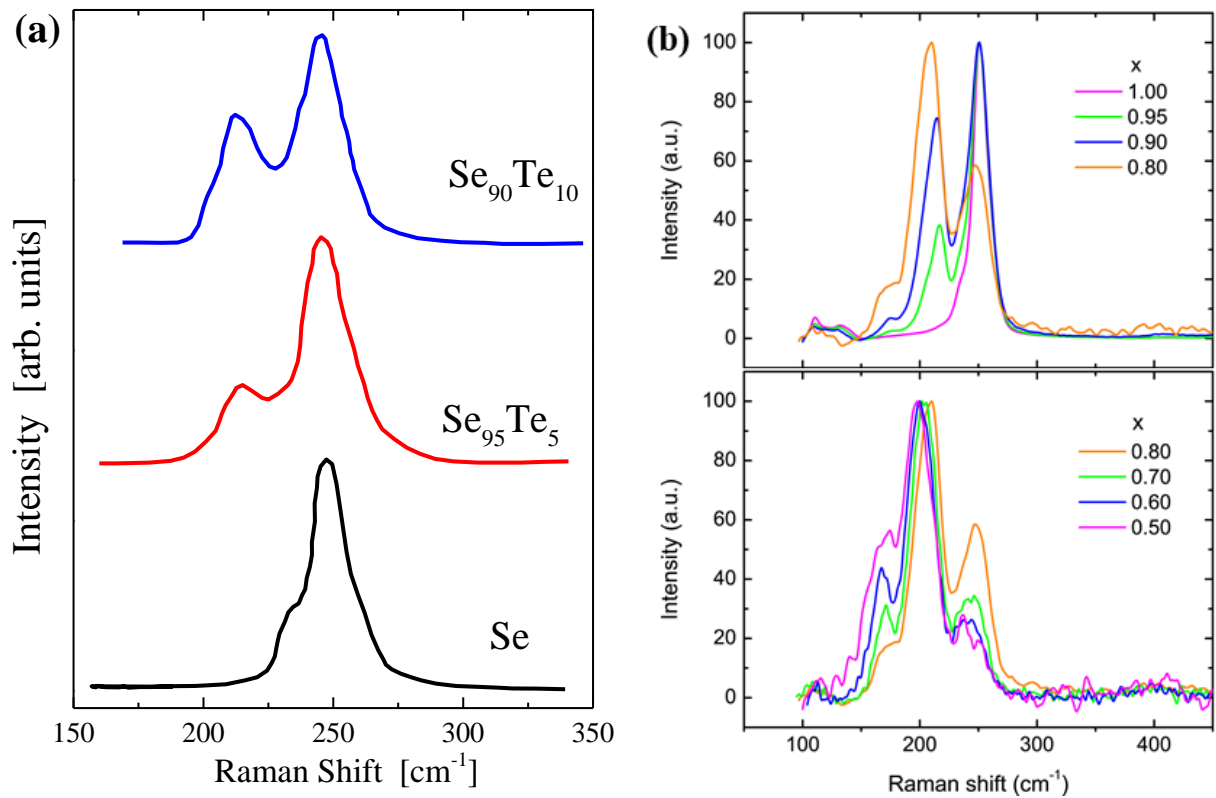

**Fig. 20:** (**a**) Raman spectra of Se-rich binary Se-Te glasses. "Adapted with permission from (Molecular structure of dilute vitreous selenium-sulfur and selenium-tellurium alloys, A. T. Ward, J. Phys. Chem. (1970) **74**(23), 4110-4115)." *Copyright © (1970), American Chemical Society*. (**b**) Raman spectra of binary Se$_x$Te$_{1-x}$ glasses; *top*: $0.8 \leq x \leq 1.0$, *bottom*: $0.5 \leq x \leq 0.8$. Reprinted from Ref. [133]. © *2018 The American Ceramic Society*.

In the most recent study, Tverjanovich *et al.* [133] investigated by Raman scattering and DFT calculations Se$_x$Te$_{1-x}$ bulk glasses in the composition range $0.5 \leq x \leq 1.0$. Figure 20(b) shows representative Raman spectra of these binary glasses. The spectra show a systematic change of the relative intensity between the stretching modes of all pairs of atoms, i.e. Se–Se (250 cm$^{-1}$), Se–Te (210 cm$^{-1}$) and Te–Te (175 cm$^{-1}$). DFT calculations of various small-sized clusters were interpreted as indicating a weak preference of a random distribution of the Se–Se, Se–Te, and Te–Te chemical bonds with some preference for heteronuclear bonding.

## 9. High pressure Raman studies

### 9.1 Elemental Sulfur

Glassy sulfur is metastable at ambient conditions. Hence the Raman studies of sulfur pertain to the study of the orthorhombic phase, $\alpha$-S$_8$, and it modifications upon pressure including its amorphization. Zallen [134] reported the first study of elemental crystalline sulfur by Raman scattering over a relatively low pressure range, i.e. 10 kbar (or 1 GPa), under hydrostatic conditions. This study showed that the pressure-induced shift of the vibrational frequencies in such a molecular crystal is strikingly inconsistent with the frequency-volume Gruneisen relation. The Gruneisen parameter was found to sharply decrease upon increasing the mode frequency.

Several high-pressure Raman studies which followed provided a rather unclear view of the various transitions that $\alpha$-$S_8$ undergoes at elevated pressures as pointed out in [135] in which a compilation of previous studies has been presented. The main reason is that photo- and/or thermal-induced phenomena, triggered by laser radiation, can take place at elevated pressures where the bandgap of sulfur decreases drastically [136] and becomes comparable to the excitation wavelength. Near-resonance conditions can critically affect the phase transitions caused by pressure. Pressure-induced amorphization was observed in a narrow pressure interval, although the amorphization was only partial and reversible [135]. The crystal-to-amorphous transition appears at pressure values of 3-4, 5-6, 10, and 12-13 GPa for excitation wavelengths 488, 514.5, 600, and 632.8 nm, respectively [135]. The transition pressure depends on laser wavelength, i.e. it increases when increasing the wavelength of the radiation. Photoamorphization has been considered as a photochemical reaction dependent upon pressure. The reaction considers the excitation of an electron to an antibonding state and its consequent recombination to a chain-like configuration instead of the structurally similar ring (*cis-trans* isomerization). The energy required to enable the isomerization change is provided by the incident photons. It is therefore expected, and indeed verified experimentally, that the transition is not observed even up to very high pressures using FTIR spectroscopy up to 10 GPa [137] and FT-Raman spectroscopy up to 5.6 GPa 138].

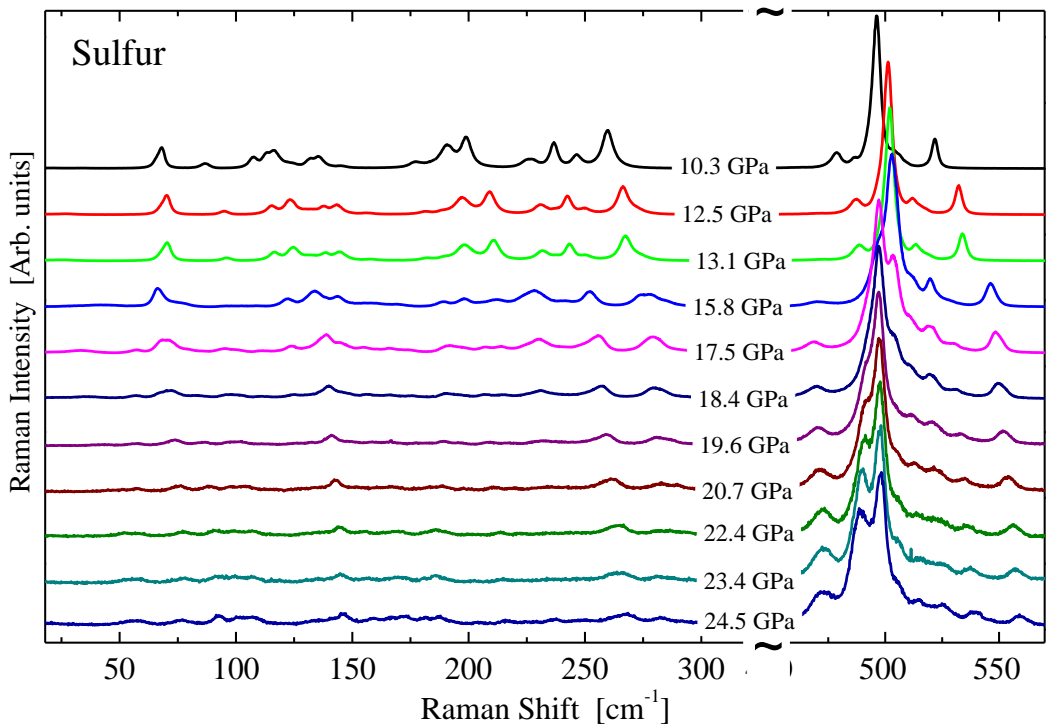

**Fig. 21:** High pressure Raman spectra of elemental sulfur for pressures from 10.3 GPa up to 24.5 GPa. Adapted from Ref. [139].

In a more recent detailed high-pressure study up to 24.5 GPa using the 752.5 nm excitation source and a low power density on the sample (150 W cm$^{-2}$) evidence has been provided against any structural transitions or photoamorphization of elemental sulfur [139]. Typical high-pressure spectra are shown in Fig. 21. Even up to the very high pressures of that work, no significant Raman peak broadening was observed showing that with proper use of wavelength and laser power, purely pressure-induced changes with rather negligible contribution from photo- or thermal effects are observed [139]. These findings and conclusions are in accordance X-ray diffraction experiments [140] which explored structural effects conducted under the combined effect of pressure and temperature.

### 9.2 Glassy Selenium

Quite a few attempts have been focused on the study of the pressure-induced structural changes in amorphous and glassy Se, whereas the crystalline polymorphs have been studied in more detail by Raman scattering [141]. An early study of amorphous Se combining X-ray diffraction and Raman scattering was interpreted to provide support for the onset of an amorphous to crystalline transition at 11 GPa [142]. While this conclusion was supported by the diffraction data, the Raman spectra are essentially inconclusive, since the material underwent crystallization even at the starting pressure, 0.5 GPa, owing to either the laser wavelength used or the laser fluence. Yang *et al*. [143] reported the first detailed Raman study of the pressure-induced structural changes of glassy Se. The data showed that the material remains glassy up to a pressure of ~10 GPa where it turns to a crystalline form. By analyzing the new Raman bands above the crystallization threshold, the authors concluded that the high-pressure crystal structure is not a single hexagonal phase, but is composed of a mixture of the hexagonal phase and an unknown high-pressure phase, which XRD identified as the tetragonal phase.

In a more recent study, He *et al*. [144] studied the high-pressure behavior of glassy Se focusing in the pressure range up to 4.3 GPa. The main Raman band ~250 cm$^{-1}$ was shown to experience red-shift upon increasing pressure with a slope of ~2 cm$^{-1}$ GPa$^{-1}$. The intensity of this band was observed to decrease vs. pressure, undergoing a discontinuous change at ~2.5 GPa. The authors speculated that this sudden intensity drop could be related to an amorphous-to-amorphous transition. However, this seems highly unlikely as no significant changes in Raman bands are observed in the spectra. It should be stressed that changes in absolute Raman intensity may reflect the emergence of extrinsic factors in the experiment and cannot be considered as solid evidence for structural changes in the medium under pressure. Finally, as noted in Section 6.1, another Raman study of high-pressure structural changes of Se [86] has presented dubious conclusions as it is based on a flawed interpretation of the main Raman bands of non-crystalline Se.

## 10. Photoinduced phenomena

### 10.1 *Glassy and liquid Sulfur*

As the stable phase of sulfur at ambient conditions is the orthorhombic crystal, very limited information about the photoinduced changes of the glass are known, whereas no Raman study has been reported. Elliott [145] observed photoinduced ESR in glassy sulfur after illumination at 77 K using light near-bandgap energy, 3.1 eV. Using molecular orbital calculations, a simple dangling bond was proposed as a possible center for the ESR signal. Sakaguchi and Tamura explored systematically photoinduced effects of liquid sulfur [146,147]. Measuring the transient absorption spectra, it was suggested that polymerization may take place at temperatures below $T_\lambda$ if the $S_8$ liquid is illuminated with proper laser light (355 nm). Depending upon the laser fluence, the process was considered to occur at two stages [146]. During the first step (low fluence) and intermediate product between the ring and the chain is formed. In the second stage, the formation of polymeric chains takes place. Later, a third process for even higher fluencies was observed, where a macroscopic iridescent pattern was considered to arise from a macroscopically ordered pattern of polymeric chains [147]. First principles molecular dynamics simulations were employed to understand the photoinduced polymerization process below the transition [148]. It was found that the scission of a bond in an $S_8$ ring forms an $S_8$ chain which after ceasing electronic excitation does not lead to the reconstruction of the same ring but generates a tadpole structure where an $S_7$ ring is branched by an additional S atom. The proximity of $S_8$ chains and tadpole structures facilitate the polymerization process.

### 10.2 Glassy and amorphous selenium

Photocrystallization is an effect which has always been suspicious about its true origin, as thermal effects may intervene (unavoidably) upon illumination of a semiconductor, especially when the light energy is comparable to the optical bandgap. Therefore, care should be exercised when the term photocrystallization is used to imply a purely athermal effect [149]. Based on experimental observations (mainly done in early sixties by Soviet researchers) Dresner and Stringfellow studied in detail the kinetics of photocrystallization of Se and provided the basis to understand the effect [150]. Crystallization kinetics was quantified by monitoring the growth rate of the diameter of spherulites upon irradiation with a 100 W Hg lamp. The proposed mechanism considers the generation of hole-electron pairs. Holes diffuse and become trapped at the crystal boundary, hence determining the growth rate of the crystal. The energy released after the annihilation of an electron-hole pair is expended for the reorientation of the Se chains, leading to an arrangement similar to the crystalline configuration. The authors associated the motion of a hole with a moving broken bond.

Several studies have been focused on measuring optical anisotropy of amorphous Se films upon irradiation [151] while only a few works have dealt with a Raman scattering study of the effect [152,153]. Poborschii *et al*. [152] studied how laser light polarization affect the crystallization of a-Se. They observed that the c-axis of t-Se attains orientation perpendicular to the direction of the polarization of the incident light. Apart from the optical origin of the effect, contribution of thermal effects was not precluded. Roy *et al*. [153] investigated the influence of the light wavelength on the photocrystallization of a-Se using above- and sub-bandgap lasers. They observed that photo-crystallization was suppressed when the film of a-Se was simultaneously exposed to two light sources with photon energies astride the absorption edge. Figure 22 shows the time evolution of the Raman spectra during the photocrystallization process of a-Se films. The use of a sub-bandgap laser in Fig. 22(a) reveals enhanced crystallization as the 234-237 cm$^{-1}$ doublet grows appreciably with irradiation time. The simultaneous exposure of the scattering volume to an above-bandgap laser in Fig. 22(b) retards the photocrystallization effect and diminishes its magnitude. The photocrystallization suppression was found to depend on the mutual polarization of the two laser beams, being more prominent when the two beams have parallel polarizations.

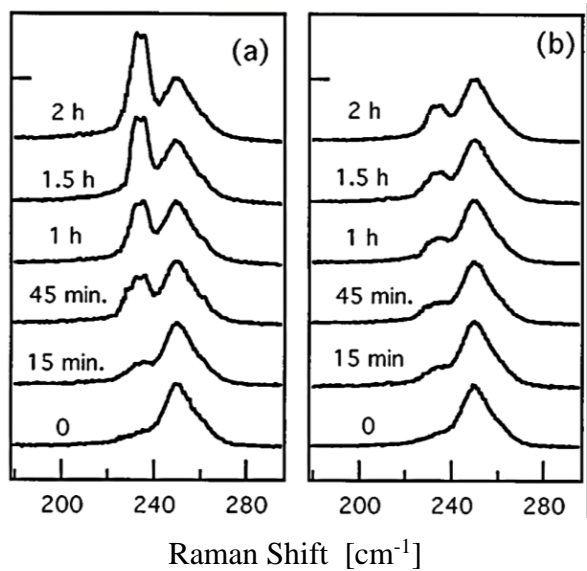

**Fig. 22:** Time dependence of the Raman spectra of a-Se upon photocrystallization. **(a)** irradiation by the 676 nm light (400 W cm$^{-2}$). **(b)** Simultaneous irradiation by the 676 nm laser as in (a) and 488 nm (200 mW cm$^{-2}$). Adapted from Ref. [153]. *Rights managed by AIP Publishing.*

A detailed investigation of photo-crystallization of amorphous Se films was presented by Tallman et al. [154] using the 632.8 nm radiation. These authors studied amorphous 15 μm-thick a-Se films sandwiched between CeO$_2$ and Sb$_2$S$_3$ (< 100 nm) films, which is the structure used to prepare high-gain avalanche rushing photoconductor devices. The real temperature of the scattering volume, $T_{loc}$, was monitored by the Stoke – anti-Stokes intensity ratio. The crystallization fraction $f_c$ was defined as the intensity ratio of the combined 234-237 cm$^{-1}$ doublet over the total intensity in the 210-280 cm$^{-1}$ spectral area. It was shown that using a flux of 17 W cm$^{-2}$ at 277 K, it takes almost 7.5 h for complete crystallization. Photo-crystallization was not observed at two temperature intervals: (a) for $T < 260$ K and (b) for a 15 K wide range around $T_g$ (~310 K). The temperature of the irradiated volume seems to appreciably influence the onset time of photo-crystallization defined as the time needed to achieve $f_c \geq 0.05$, i.e. 5% crystallization of the amorphous structure. Figure 23(a) shows the Raman spectra at the saturated state after irradiation for the designated temperatures of the scattering volume, which were obtained after a period of time denoted besides the spectra. The spectra reveal that the sample temperature determines the extent and time-scale of photo-crystallization [154]. The onset time depends upon temperature in the fashion shown in Fig. 23(b). This non-systematic dependence of the extent and kinetics of photo-crystallization against the sample temperature was accounted for by invoking arguments which consider the effects of local strain on the secondary growth of crystalline nuclei in a-Se [154]. Strain relaxation near $T_g$ was anticipated to prevent crystal growth. It should be noted here that the criterion employed to determine the onset time ($f_c \geq 0.05$) might be perplexing, since a crystallinity of this order intrinsically accompanies glassy or amorphous selenium [79].

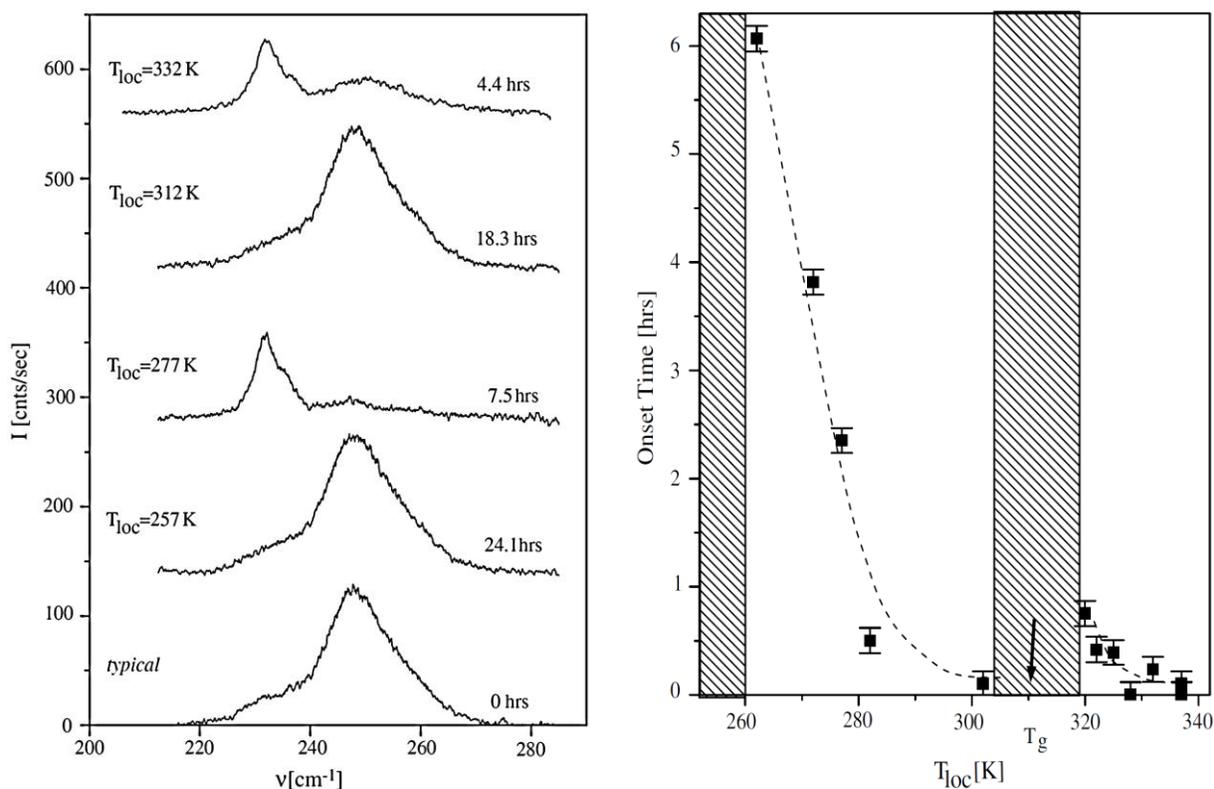

**Fig. 23:** **(a)** Raman spectra of irradiated a-Se films (17 W cm$^{-2}$) after the system has reached saturation of photo-crystallization at different temperatures ($T_{loc}$). **(b)** Variation of the irradiation time until onset of photo-crystallization at various temperatures. Adapted from Ref. [154]. *Copyright © 2008 Elsevier B.V. All rights reserved.*

Closing this section, it is worth mentioning that both t-Se and a-Se films when irradiated by low intensity sub-bandgap light at low temperature (77 K) show morphological changes (mass transport) while t-Se exhibits at the same time amorphization. The effect was termed as photomelting [155], while it was shown to be purely athermal. The term "melting" might sound improper as not temperature rise takes place. The authors suggested that photocrystallization is a combined process of melting and crystallization being athermal and thermal, respectively.

### 10.3    *Photo-oxidation and photo-amorphization of t-Te*

Information on amorphous tellurium has been meager as discussed in previous section, owing to the difficulty in obtaining and maintaining this element in the non-crystalline phase. However, owing to the low energy of the Te−Te bond, elemental Te exhibits strong tendency towards oxidation especially under illumination. In contrast to sulfur, which energetically prefers the *cis* conformation, the chain-like structure of t-Te (hexagonal packing of helical chains) eventually determines its tendency to grow in anisotropic 1-D nanostructures. Vasileiadis *et al*. [97] reported a laser-assisted method for the controlled fabrication of various Te nanostructures such as nanorods and nanotubes. Controlled illumination of such structures can lead to morphology and phase transformations (photo-oxidation) to nanowires with a core of Te and a sheath of $TeO_2$; the sheath thickness can be controlled to obtain entirely $TeO_2$ nanowires. A collection of field-emission scanning electron images illustrating the photoinduced process taking place is shown in Fig. 24.

Raman spectra, using the 441.6 nm laser wavelength as excitation source, provided information that apart from the photo-oxidation process, photo-amorphization of t-Te proceeds in parallel, [98]. Therefore, *in situ* Raman scattering was employed to observe and quantify the kinetics of both processes against light fluence. Characteristic spectra of t-Te prior to and after irradiation, bulk glassy TeO₂ and crystalline TeO₂ are sown in Fig. 25(a). Representative time resolved Raman spectra upon irradiation are shown in Fig. 25(b). The Raman bands of t-Te at low energy ($< 150$ cm$^{-1}$) becomes at early stages of the irradiation masked by the overwhelming increase of the amorphous a-Te band at ~165 cm$^{-1}$ as indicated by the vertical dashed line. In parallel, Raman bands of amorphous TeO₂ appear at high energies ($>300$ cm$^{-1}$). The analysis showed that amorphization exhibits faster kinetics upon increasing the light fluence, whereas the time scale of the photo-oxidation process remained unaffected. High-resolution transmission electron microscopy provided images about the interphase of t-Te and TeO₂, showing that an ultrathin a-Te layer is sandwiched among them. Lowering the laser fluence by almost two orders of magnitude (i.e. $10^2$ W cm$^{-2}$) the photoinduced process changed drastically. In this case, t-Te is led directly to the crystalline oxide phase avoiding the formation of amorphous layers of Te and TeO₂. A series of Raman spectra illustrating this peculiar effect are presented in Fig. 25(c). This finding indicated the formation of a sharp interphase between the two crystalline phases. As has been pointed out in [98] laser-assisted nano-structuring of elemental Te is facilitated by its high tendency towards crystallization. Selenium does not show potential for exhibiting analogous effects. Telluride materials, such as the phase change memory alloys, are also poor glass-formers. As recent trends show, focus has been directed to miniaturizing the memory cell size. In this context, laser-assisted nanostructuring might be central for obtaining materials with significantly improved functionalities.

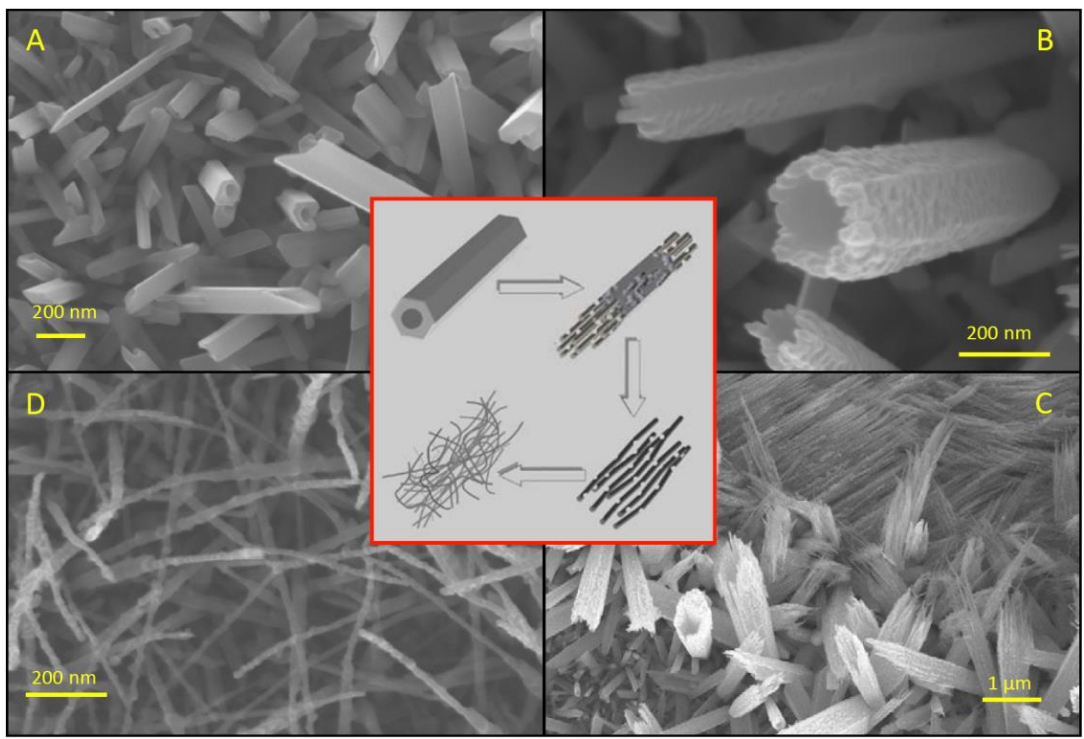

**Fig. 24:** Field-emission scanning electron images showing the light-induced transformation of t-Te nanotubes (A) to ultrathin nanowires (D) which can be either α-TeO₂ and/or Te-core/TeO₂-sheath nanostructures with a size distribution in the range 10-20 nm. (B) and (D) shows intermediate steps of the nanostructuring of the nanotube surface due to photo-induced oxidation. The schematic inset shows the steps of the photo-induced mechanism. Adapted by permission: [Springer Nature] [**Sci. Rep.** (Laser-assisted growth of t-Te nanotubes and their controlled photo-induced unzipping to ultrathin core-Te/sheath-TeO₂ nanowires, **3**: art. no.1209 (2013)].

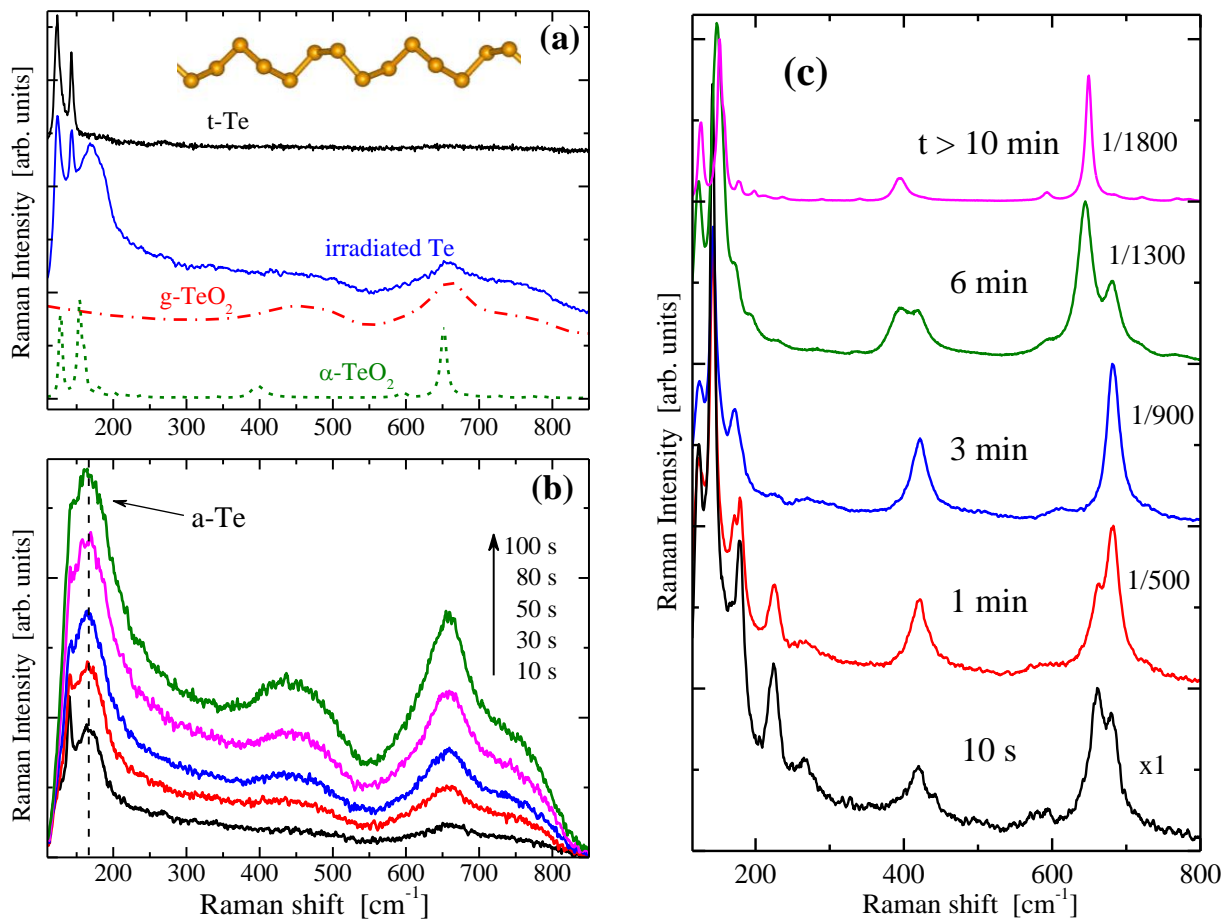

**Fig. 25:** **(a)** Raman spectra of various Te and TeO₂ phases. **(b)** Time resolved Raman spectra of t-Te irradiated with $10^4$ W cm⁻² recorded at short time intervals. **(c)** Kinetics of photoinduced surface growth of crystalline TeO₂ avoiding the formation of an amorphous oxide product ($10^2$ W cm⁻²). The magnification factors are shown for each spectrum. Adapted from Ref. [**Error! Bookmark not defined.**]. *Rights managed by AIP Publishing.*

## 11. Concluding Remarks

In the preceding review, a wide variety of information about structure and properties of the elemental chalcogens have been brought together. The focus has primarily been to collect and assess results inferred by Raman scattering on all non-crystalline phases of S, Se and Te. This survey has made evident that besides furnishing a wealth of structural information, Raman scattering lends itself to the study of a number transitions occurring under the influence of some external stimulus, such as temperature, pressure, irradiation and so on, thus offering useful thermodynamic information. It has also become clear that despite their monoatomic nature, elemental chalcogens display a rich diversity of Raman spectra whose precise interpretation is the prerequisite to understand the structure of binary and multicomponent chalcogenide glasses.

The poor stability of sulfur glass has been the cause for the sporadic studies and scarce attention it has received. The interest has been directed to studies of the liquid phase and the rich variety of implications accompanying the polymerization or $\lambda-$transition. However, the glassy state of quenched sulfur deserves heightened interest as it exhibits unique phenomena, far from being adequately understood. Glassy Se has undoubtedly been the most thoroughly explored chalcogen in its glassy state and has since early studies met important applications. The glass structure, in particular the ring-to-chain content, is still a source of controversy. This monoatomic amorphous solid, obeying two-fold coordination, has been frequently used as paradigm of photo-induced structural changes, giving rise to some blatant, though not always realistic, models of the atomic arrangements in photostructural changes. Despite that tellurium-based phase-change alloys have received tremendous attention, the element has mostly been marginalized owing to its failure to form bulk glass and the very poor stability of the amorphous film. However, a large number of studies have been dedicated to the structural study of the liquid state in order to parallelize this information to understand structure of the disordered solid. Notably, Te is the most liable to light chalcogen as it can be restructured at the nanoscale, being also capable of experience photo-amorphization and photo-oxidation providing a number of useful structures and interfaces.

Structure and phase transitions in elemental chalcogens is a long-standing field of research. Dealing with century-old issues that have not yet reached consensus, is an arduous task. The current review is not meant to be a complete collection of results on Raman scattering of elemental chalcogens. Based on certain criteria, we have instead aimed at providing a critical assessment of selected reports attempting to provide a balanced discussion of the relevant literature, stressing at specific cases weaknesses and confusion caused by misinterpreting Raman spectra. Many more papers than those referenced herein have been published. Although other important reports may have been non-deliberately omitted, our intention was not to cite all the published articles, but to make reference to those that seemed of worth to be discussed in this short review, for their originality at the time they were published and for the rational analysis of the Raman spectra.

## Acknowledgments

I want to thank a number of colleagues whose collaboration over the past several years has been instrumental in conducting systematic research on elemental chalcogens. Therefore, I wish to acknowledge collaboration to Dr. K. S. Andrikopoulos, Dr. D. Th. Kastrissios, Dr. A. Kalampounias, Prof. A. Chrissanthopoulos and Prof. G. N. Papatheodorou. Finally, I want to thank Ms. M. Perivolari for managing the copyrighted material presented herein.

This work was supported by the project "National Infrastructure in Nanotechnology, Advanced Materials and Micro-/ Nanoelectronics" (MIS 5002772) which is implemented under the Action "Reinforcement of the Research and Innovation Infrastructure", funded by the Operational Programme "Competitiveness, Entrepreneurship and Innovation" (NSRF 2014-2020) and co-financed by Greece and the European Union (European Regional Development Fund).